\pdfoutput=1
\documentclass[nofootinbib,a4paper,longbibliography,groupedaddress,amsfonts,amssymb,amsmath, twocolumn]{revtex4-2}

\usepackage[T1]{fontenc}
\usepackage[utf8]{inputenc}
\usepackage[most]{tcolorbox}

\usepackage{lmodern,microtype,changepage,amsthm,mathtools,bm,tabularx,graphicx,enumitem,booktabs,tikz,xcolor,tabularray,mathtools,physics}
\usepackage[scr=boondoxo,bb=boondox]{mathalfa}

\definecolor{darkblue}{rgb}{0.1,0.1,.7}
\usepackage[colorlinks, linkcolor=darkblue, citecolor=darkblue, urlcolor=darkblue,
linktocpage, breaklinks, bookmarksnumbered,
bookmarksdepth=2]{hyperref}
\usepackage{orcidlink}

\newcommand{\SB}{S_{\mathrm{B}}}
\newcommand{\SE}{S_{\mathrm{E}}}
\newcommand{\SIR}{S_{\mathrm{IR}}}
\newcommand{\PsiUV}{\Psi_{\mathrm{UV}}}
\newcommand{\PsiIR}{\Psi_{\mathrm{IR}}}
\newcommand{\PiT}{\Pi^{\perp}}
\newcommand{\PiL}{\Pi^{\parallel}}
\newcommand{\dhat}{\hat{\delta}}

\def\<#1\>{\expval{#1}}

\renewcommand \.   {\cdot}
\newcommand   \x   {\times}

\newcommand   \half{\frac 1 2}
\newcommand   \lptl{\raise .8ex\hbox{$^\leftarrow$} \hspace{-9pt} \partial} % derivatives with arrows
\newcommand   \lrptl{\raise .8ex\hbox{$^\leftrightarrow$} \hspace{-9pt} \partial} % derivatives with arrows
\newcommand \nn {\nonumber}
\def\be#1\ee{\begin{equation}\begin{aligned}#1\end{aligned}\end{equation}}
\makeatletter
\DeclareRobustCommand\bea{\@ifnextchar[{\@@bea}{\@bea}}
\def\@@bea[#1]#2\eea{\begin{subequations}\begin{align}#2\end{align}\label{#1}\end{subequations}}
\def\@bea#1\eea{\begin{subequations}\begin{align}#1\end{align}\end{subequations}}
\makeatother
\mathcode`l="8000
\begingroup
\makeatletter
\lccode`\~=`\l
\DeclareMathSymbol{\lsb@l}{\mathalpha}{letters}{`l}
\lowercase{\gdef~{\ifnum\the\mathgroup=\m@ne \ell \else \lsb@l \fi}}%
\endgroup
\renewcommand \a  {\alpha}
\renewcommand \b  {\beta}
\newcommand   \g  {\gamma}

\newcommand   \de {\delta}

\newcommand   \e  {\epsilon}
\renewcommand \l  {\lambda}

\newcommand   \f  {\phi}

\newcommand   \s  {\sigma}
\newcommand   \w  {\omega}

\newcommand   \ka {\kappa}

\newcommand   \SU    {\mathrm{SU}}

\newcommand   \Sym   {\mathrm{Sym}}

\newcommand   \cA {\mathcal{A}}

\newcommand   \cD {\mathcal{D}}

\newcommand   \cG {\mathcal{G}}
\newcommand   \cH {\mathcal{H}}

\newcommand   \cJ {\mathcal{J}}
\newcommand   \cK {\mathcal{K}}

\newcommand   \cM {\mathcal{M}}

\newcommand   \cO {\mathcal{O}}
\newcommand   \cP {\mathcal{P}}

\newcommand   \cV {\mathcal{V}}
\newcommand   \cW {\mathcal{W}}

\makeatletter
\def\l@subsubsection#1#2{}
\makeatother

\begin{document}
\title{Holographic Wilsonian RG of Yang--Mills theory in AdS}
\author{Soner Albayrak\,\orcidlink{0000-0002-9223-0115}}
\affiliation{Department of Physics, Middle East Technical University, Ankara, Türkiye}

\begin{abstract}
In this work, we extend the equilibrium-evolution duality between stochastic quantization and Wilsonian renormalization to non-abelian gauge theories in anti-de Sitter space, for which the radial coordinate of AdS plays the role of the stochastic time. We show, to all orders in the fields for bulk theories without explicit radial dependence, that the Langevin distribution at time $t$ is the product of the ultraviolet and infrared radial wavefunctions at the cutoff $\epsilon=t$. We therefore identify this Langevin process as the ground-state (Doob) transform of the radial evolution. By explicit calculation, we also extend the existing dictionary in the stochastic-renormalization duality: the radial component of the bulk field is Zwanziger's stochastic gauge-fixing function; radial gauge is Parisi--Wu quantization; and the bulk gauge $A_z=\a^{-1}\partial_iA_i$ is Zwanziger's gauge. We also obtain the expressions for the triple-trace and quadruple-trace couplings of the holographic Wilsonian action at finite cutoff in closed form; in the infrared, they reduce to the known gluon correlators and, at the total-energy pole, to flat-space Yang--Mills amplitudes as expected. Although our explicit results are provided at tree-level and derived in AdS$_4$, we also comment on the de Sitter counterpart of these statements and on  the loop corrections.
\end{abstract}
\date{\today}
\maketitle
%\pacs{}
%\keywords{}
\tableofcontents

\section{Introduction}
The holographic duality with its most concrete realization via AdS/CFT correspondence \cite{Maldacena:1997re,Witten:1998qj} has been of great importance in contemporary theoretical physics for the past few decades. It relates a gravitational theory in a $(d+1)$-dimensional bulk to a $d$-dimensional boundary theory; and it has been extensively studied with a variety of approaches, from more traditional position \& Mellin space techniques \cite{Freedman:1998tz,Liu:1998ty,DHoker:1999kzh,DHoker:1999mqo,Mack:2009mi,Penedones:2010ue,Fitzpatrick:2011ia,Paulos:2011ie,DHoker:2002nbb,Penedones:2016voo} and boundary CFT perspectives \cite{Heemskerk:2009pn,El-Showk:2011yvt,Fitzpatrick:2012yx,Komargodski:2012ek,Aharony:2016dwx,Caron-Huot:2017vep,Alday:2017vkk,Alday:2021odx,Poland:2018epd,Bissi:2022mrs} to more recently appreciated explicit bulk analysis with technologies borrowed from the scattering amplitudes program \cite{Raju:2010by,Raju:2011mp,Maldacena:2011nz,Raju:2012zr,Arkani-Hamed:2017fdk,Arkani-Hamed:2018kmz,Farrow:2018yni,Goodhew:2020ehn,Meltzer:2020qbr,Armstrong:2020woi,Armstrong:2022mfr,Baumann:2022jpr,Benincasa:2022gtd}. This paper is a natural continuation of the recent progress in the last approach.\footnote{For a recent review of how scattering amplitude methods are being used in the curved space literature, see \cite{Albayrak:2026auu}.}

Quite independently of these developments in the study of holography, stochastic quantization has also been an active area of research where the tentative idea is not so different from that of the holography: we have two field theories, a $(d+1)$-dimensional one in which the fields evolve diffusively in a fictitious time $t$ according to a Langevin equation (hence the stochastic nature), and a $d$-dimensional Euclidean one, whose \emph{equilibrium action} governs the late-time limit $t\to\infty$ of the former \cite{Oh:2012bx}. However, unlike the curved space holography, this \emph{duality} is more of an equilibrium-evolution relation than a geometric bulk-boundary relation. Nevertheless, the schematic resemblance did lead to a sizable interest in finding a more concrete relation between these two concepts, and the literature has seen quite an effort along this direction \cite{Minic:2010pw,Mansi:2009mz,Polyakov:2000xc,Lifschytz:2000bj}.

The first major result in the establishment of such a relation can be taken as the conjectured identification of the bulk action on one side with the Fokker-Planck action on the other \cite{Mansi:2009mz}, which is roughly the identification of the partition functions of the holographic and stochastic models in the classical limit. This was later sharpened into a dictionary between the radial flow of the holographic Wilsonian effective action \cite{Heemskerk:2010hk,Faulkner:2010jy} and stochastic time evolution \cite{Oh:2012bx}: the stochastic time is the radial coordinate, the Fokker--Planck Hamiltonian is the radial Hamiltonian, and the multi-trace couplings of the Wilsonian action are fixed by stochastic correlation functions.

This dictionary has been worked out in a growing number of examples\footnote{\label{footnote: cases that has been analyzed} \emph{At the free level:} free scalars and $U(1)$ gauge fields \cite{Oh:2012bx}; conformally coupled scalars \cite{Jatkar:2013uga}; massless and massive fermions \cite{Oh:2013tsa,Moon:2017btx}; scalars of arbitrary mass \cite{Oh:2015xva}. \emph{Beyond the free level:} the conformally coupled $\f^3$ theory in AdS$_6$ \cite{Oh:2021bxx}; the conformally coupled $\f^4$ theory in AdS$_4$ \cite{Lee:2023ynb}; massive scalars with self-interactions and multi-trace deformations \cite{Kim:2023bhp,Chae:2024zyn}.} and the conceptual grasp around it has certainly improved;\footnote{For instance, the dictionary is expected to extend to all multi-trace couplings of the conformally coupled theory \cite{Oh:2021bxx}, and Euclidean actions with explicit dependence on the stochastic time have already been explored \cite{Chae:2024zyn}. See also \cite{Kim:2021rik,Shin:2026xoo,Kamei:2024iei,dosReis:2018xxr,Sathiapalan:2017frk,Dharanipragada:2020ibl} for other related developments. } however, no analysis of gauge fields beyond the free level has been conducted yet,\footnote{The only available work so far to our knowledge is the free level $U(1)$ analysis of \cite{Oh:2012bx}, which restricts to the transverse modes and recovers the bulk Lagrangian only up to gauge degrees of freedom.} despite it being an important gap in the literature.\footnote{ Besides its physical relevance, stochastic analysis of non-abelian gauge theories has a historical importance as well because non-abelian gauge theories were among the main motivations for stochastic quantization in the first place \cite{Parisi:1980ys,Zwanziger:1981kg,Baulieu:1981ec}. Additionally, its noiseless limit, the Yang--Mills gradient flow \cite{Narayanan:2006rf,Luscher:2010iy}, has since been related to renormalization-group transformations \cite{Carosso:2019qpb,Sonoda:2020vut} and to an emergent holographic direction \cite{Aoki:2015dla}.} We will address this situation by explicitly analyzing gluon correlators in AdS$_4$, which have been shown to take remarkably simple forms if written in a judiciously chosen coordinate system \cite{Albayrak:2019asr}.\footnote{Motivated by the manifestly direct comparison with the flat-space scattering amplitudes, (A)dS correlators have been analyzed for the past decade by a group of researchers (author included) in \emph{Poincar\'e patch} with radial direction in position space and boundary directions in momentum space, leading to explicit results both at tree and loop level \cite{Albayrak:2018tam,Albayrak:2019asr,Albayrak:2019yve,Albayrak:2020isk,Albayrak:2020bso,Albayrak:2020fyp,Albayrak:2023jzl,Albayrak:2023kfk,Albayrak:2024ddg}. Unexpectedly, this approach turned out to be rather fruitful and has been extensively utilized in various related fields (see \cite{Albayrak:2026auu} and the references therein).} Although \emph{some} form of a reduction in complexity is expected as four-dimensional Yang--Mills theory on AdS enjoys the classical Weyl invariance, making its tree-level perturbation expansion identical to diagrams on a flat half-space,\footnote{Not differently, this is also the reason why conformally coupled scalars have seen so much attention in the stochastic program.} the form of the Witten diagrams that we will be utilizing benefits from far more simplifications, thanks to the collaborative effort on the holography program.\footnote{Besides the more practical results due to the \emph{axial gauge} choice (such as every gluon Witten diagram factorizing into vertex factors and the corresponding diagram of a conformally coupled scalar, acted on by a simple operator), the holography program has also provided us with insight from different angles, including colour-kinematics duality and the double copy \cite{Albayrak:2020fyp,Armstrong:2020woi,Diwakar:2021juk,Zhou:2021gnu,Herderschee:2022ntr,Cheung:2022pdk,Lipstein:2023pih}, on-shell recursion and soft limits \cite{Mei:2024abu,Chowdhury:2024wwe,Albayrak:2024ddg,Pathak:2026nue}, the amplitude expansion of gluon correlators \cite{Gomez:2026yno}, and self-dual Yang--Mills theory in AdS$_4$ \cite{Heuveline:2026knw}.}

In this paper we combine these developments. We extend the stochastic--holographic dictionary to $\SU(N)$ Yang--Mills theory in AdS$_4$ at tree level and use it to analyze the gauge sector of the holographic Wilsonian action. Our main points are as follows:

\paragraph{Holographic realization of stochastic gauge fixing:} The $(d+1)$-dimensional formulation of stochastic gauge fixing of Baulieu and Zwanziger \cite{Baulieu:1999wz} is realized by the bulk gauge field: the radial component $A_z$ is Zwanziger's gauge-fixing field, radial gauge $A_z=0$ is the Parisi--Wu process, and other bulk gauges are other stochastic gauge fixings.

\paragraph{Closed-form multi-trace couplings of gluons at finite cutoff:} By utilizing the operator of \cite{Albayrak:2019asr} that maps scalar to gluon diagrams as the functional calculus of the stochastic drift, we can construct a closed-form map between the triple-trace \& quadruple-trace flows of gluons and scalar results.\footnote{In radial gauge, infrared limit of multi-trace flows of gluons reproduces the colour-ordered gluon correlators of \cite{Albayrak:2019asr} and, at the total-energy pole, flat-space Yang--Mills amplitudes.} We also provide closed form results for the exchange seed at finite cutoff and its longitudinal limit.

\paragraph{Analysis of gauge dependence in Yang--Mills:} We show that, on transverse data, the entire gauge dependence of the quadruple-trace couplings is in a single longitudinal zero mode, which is absent when the Gauss law is imposed and is inconsequential in the flat-space limit.\footnote{Explicitly, we show in \S~{\ref{sec: gauge dependence}} that in the longitudinal channel, the quadruple-trace flow is the Coulomb energy of the radial colour charge minus a zero mode fixed by the gauge condition, and the zero mode carries the entire gauge dependence. In radial gauge it is the zero mode of a Brownian bridge (Brownian motion on $0<z<\e$ conditioned to vanish at both ends, whose covariance $z_<(\e-z_>)/\e$ is the longitudinal propagator of the $0<z<\epsilon$ AdS slab) and decays as the inverse cutoff. This is the origin of the power-law relaxation of the gauge-mode parts. In Zwanziger's gauge it relaxes exponentially to an $\a$-dependent value, and it is absent when the Gauss law is imposed. It is a redundant coupling, which a gauge rotation of the cutoff data removes. It has no total-energy pole, so the flat-space limit is gauge independent. Its value in Zwanziger's gauge also follows from the stationary Fokker--Planck equation, without reference to the bulk, which provides a non-trivial consistency check of our quartic formulae for a drift that is not a gradient.}

\paragraph{Further remarks:} We also comment on the Chern--Simons solution of the zero-energy Hamilton--Jacobi equation (\S~{\ref{sec: CS branch}}), on the form of the loop-level statement (\S~{\ref{sec: proposition}}, \S~{\ref{sec: outlook}}), and on the de Sitter counterpart of our results (\S~{\ref{sec: dS reading}}).

The paper is organized as follows. In \S~{\ref{sec: reviews}} we review stochastic quantization of gauge fields, the holographic Wilsonian dictionary, and the Yang--Mills theory in AdS$_4$. We then proceed to \S~{\ref{sec: main tools}} where we recast the relation of \cite{Albayrak:2019asr} as a functional calculus and present the all-order tree-level argument with which we derive the gauge-sector dictionary in \S~{\ref{sec: gauge sector}} and the explicit multi-trace flows and their gauge dependence in \S~{\ref{sec: flows}}. We discuss our results in the context of cosmology in \S~{\ref{sec: dS reading}} and conclude with a brief discussion of open directions in \S~{\ref{sec: outlook}}. The appendices collect several conceptual and technical details and the attached \texttt{Mathematica} file re-derives the formulae of sections~\ref{sec: main tools}--\ref{sec: flows} from their definitions (the gluon statements mechanically from the Lagrangian) and contains every check quoted in the text.

\section{Reviews}
\label{sec: reviews}
\subsection{Stochastic quantization}
\subsubsection{Basics}
Although many excellent discussions of stochastic quantization exist elsewhere,\footnote{One standard review is \cite{Damgaard:1987rr} for essential details. We also recommend \cite{Mansi:2009mz} whose appendix we will partially follow in this section.}$^{,}$\footnote{For a more historical discussion of why stochastic quantization was proposed as an alternative quantization scheme, we refer to \cite{Parisi:1980ys} (\emph{bypassing gauge-fixing and the BRST formalism in non-abelian gauge theories}), \cite{Baulieu:1981ec} (\emph{establishing the equivalence with the Faddeev--Popov ansatz}), \cite{Zwanziger:1981kg} (\emph{avoiding the Gribov ambiguity}), \cite{Marchesini:1981ry} (\emph{obtaining the loop equations of gauge theories as the equilibrium condition}), \cite{Parisi:1980nh, Parisi:1981xz} (\emph{enabling better computer simulations}), and \cite{Alfaro:1982hq,Greensite:1982mf,Guha:1982uj} (\emph{understanding quenched reduced systems better}). For further details, see also \cite{Gozzi:1983qxk,Zinn-Justin:1986nph}. } we would like to quickly review the fundamentals for the sake of completeness.

Consider the generalized Stokes theorem $\int_{\cM}d\Omega=\int_{\partial\cM}\Omega$ for any $d$-form $\Omega$, and propose a  generalization of it in field theory \cite{Baulieu:1988jw}
\be
S_d[\f]=\int_{\partial \cM}\Omega[\f(x)]\;,\quad S_{d+1}[\f]=\int_{ \cM}d\Omega[\f(x)]
\ee
where we take the actions $S_d[\f]$ and $S_{d+1}[\f]$ to be schematically related to each other as
\be
S_{d+1}[\f]=\int_{\cM} dt\wedge \frac{d\phi}{d t}\frac{\delta S_{d}[\f]}{\de\f}\;.
\ee
for the additional dimension being parameterized as  $t$, a \emph{fictitious} time parameter. The invariance of the action against how the field is extended into the bulk is ensured by the existence of a topological invariant via the constraint $S_{d+1}[\f+\delta \f]=S_{d+1}[\f]$ whenever $\de\f\evaluated_{\partial\cM}=0$. This \emph{symmetry} can be fixed by a gauge choice that we conveniently choose to be the Langevin equation: the standard follow-up procedure\footnote{This roughly amounts to applying the BRST formalism by introducing ghost fields and integrating out the auxiliary field introduced by the gauge choice.} then leads to the white-noise field $\eta$ with which we identify the final form of the model as the supersymmetric generalization of the Fokker--Planck action. In this paper however, we will work with the non-supersymmetric version.\footnote{One usually recovers non-supersymmetric version from the supersymmetric one by specifying initial/boundary conditions for the field $\f$ that break the supersymmetry \cite{Gozzi:1983qxk}.}

The aforementioned Langevin equation reads as
\be
\label{eq: langevin}
\frac{\partial\f(t,x)}{\partial t}=-\ka\fdv{\SE[\f]}{\f(t,x)}+\eta(t,x)\;,
\ee
which is interpreted as evolution of the field $\f$ that quantizes a $d$-dimensional Euclidean field theory with action $\SE[\f]$. The white noise $\eta$ is Gaussian,
\be 
\<\eta(t,x)\eta(t',x')\>=2\ka\,\de(t-t')\de^d(x-x')\;,
\ee 
and it represents the coupling to a ``thermal'' reservoir.\footnote{The word thermal should not be taken too seriously as we are not really working with a finite temperature quantum field theory (i.e. no compactified direction on the manifold). We use the thermal reservoir as a reminder that we are dealing with \emph{equilibrium} dynamics on the boundary.}

On the bulk side, the Fokker-Planck action has the form\footnote{This follows from the Onsager--Machlup path integral.}
\be 
\label{eq: fokker planck action}
S_{FP}=\int dt\int d^dx\Bigg[\frac{1}{4\ka}\left(\frac{\partial \f}{\partial t}+\ka\fdv{\SE}{\f}\right)^2
-\frac{\ka}{2}\frac{\de^2 \SE[\f]}{\de \f^2}\Bigg]\;.
\ee 
and the probability distribution $P[\f,t]$ of this action does relax to $e^{-\SE}$ in the limit $t\to\infty$: this ensures that equal-time correlators of $\f$ asymptote to the Euclidean correlation functions of $\SE$ on the boundary. With a little calculation,\footnote{Set $\ka=\half$, expand the square, identify the cross term $\half\,\partial_t\f\,\de\SE/\de\f=\half\,d\SE/dt$ as a total derivative, and do a Legendre transformation: see \S~2.1.2 of \cite{Oh:2012bx} for explicit details.} we can turn the Fokker--Planck equation into a Euclidean Schr\"odinger equation,
\be 
\partial_t\psi_S=-H_{FP}\psi_S
\ee 
with
\be
\label{eq: HFP}
H_{FP}=\int d^dx\left[-\half\frac{\de^2}{\de\f^2}+\frac18\left(\fdv{\SE}{\f}\right)^2-\frac14\frac{\de^2\SE}{\de\f^2}\right]\;.
\ee
and the probability distribution $P$ identified as
\be 
\label{eq: definition of probability distribution}
P=\psi_S\,e^{-\SE/2}
\ee 

\subsubsection{Subtleties with gauge fields}
For a gauge field $A^a_i$, gauge invariance of $\SE$ implies $D_i\,\de\SE/\de A_i=0$, i.e.\ the drift is orthogonal to the gauge orbits, while the noise is not. In the Parisi--Wu formulation \cite{Parisi:1980ys}, gauge-invariant observables equilibrate, whereas the gauge modes perform an unrestricted random walk and $P$ has no normalizable equilibrium along the gauge orbits. We can however damp the gauge modes without affecting gauge-invariant observables by modifying the drift,
\be
\label{eq: Zwanziger}
\partial_tA_i=-\ka\fdv{\SE}{A_i}+D_iv+\eta_i\;,
\ee
with an arbitrary function $v[A]$, for instance $v=\a^{-1}\partial_iA_i$ \cite{Zwanziger:1981kg}. If we further choose $v$ to be \emph{time component} of $(d+1)$-dimensional gauge field, this damping can be realized in a bulk gauge covariant way \cite{Baulieu:1999wz},\footnote{The same device appears as the gauge-fixing term of the Yang--Mills gradient flow \cite{Luscher:2010iy}, and as the DeTurck trick in the mathematical literature on the stochastic quantization of Yang--Mills theory \cite{Chandra:2020dpn,Chandra:2022bkt}.} as 
\be
\label{eq: BZ}
F_{ti}\equiv\partial_tA_i-D_iA_t=-\ka\fdv{\SE}{A_i}+\eta_i\;,\qquad A_t\equiv v\;.
\ee
This motivation for the choice of $v$ will have direct relevance from the holographic perspective because we will identify $A_t$ as the radial component of the bulk gauge field.

\subsection{Holographic Wilsonian renormalization group}
\label{sec: holographic wilsonian rg}
It is straightforward to observe that the energy scale of a boundary theory corresponds to the radial coordinate of the AdS bulk \cite{Maldacena:1997re,Gubser:1998bc,Witten:1998qj,Susskind:1998dq}: next step from here is the postulation of a map between the radial cut-off in AdS and a Wilsonian cut-off in the boundary CFT \cite{Heemskerk:2010hk, Faulkner:2010jy}.\footnote{For more details, see also \cite{Skenderis:2002wp,deBoer:1999tgo,Bianchi:2001kw,Bianchi:2001de,Balasubramanian:1999jd,Alvarez:1998wr,deBoer:2000cz,Martelli:2002sp,Fukuma:2002sb,Bianchi:2000sm}.} For bulk fields $\f(z,x)$ and boundary operators $\Phi(x)$, this postulate takes the form
\be
\label{eq: holographic renormalization bulk vs boundary}
\int\cD\f\evaluated_{z>\e}e^{-S_{\text{bulk}}}={}&\int\cD\Phi\evaluated_{k<\de^{-1}}
\\&\times e^{-S_{\text{boundary}}+\int d^dx\, \f_\e(x)\cO(x)}\;,
\ee
where $\f_\e(x)\equiv\f(\e,x)$ and where $\cO(x)$ form a complete set of local single-trace operators built out of $\Phi(x)$.\footnote{Although we only consider single-trace operators in \eqref{eq: holographic renormalization bulk vs boundary}, the full bulk path integral generates general multi-trace terms. This is roughly because the complete path integral includes the integration over $z\le \e$, and carrying out that integration first generates higher-trace terms from the single-trace operators \cite{Heemskerk:2010hk,Li:2000ec,Petkou:2002bb}.}

The cut-off's in the map are the content of the postulate: from the bulk perspective, we effectively shift the boundary from $z=0$ to $z=\epsilon$; from the boundary perspective, we put a UV cut-off $1/\de(\epsilon)$ in the momentum space, which we will take as $1/\epsilon$.\footnote{Beyond pure AdS, $\de(\e)\ne\e$ and should be estimated in terms of a geodesic in the bulk which connects two boundary points and reaches down to $z=\e$ \cite{Susskind:1998dq}.} We can now define the \emph{Wilsonian boundary action}\footnote{From the bulk perspective, $\PsiUV[\f_\e;\e]$ is the integration of $e^{-S_{\text{bulk}}}$ in the UV region $0<z<\e$ at fixed $\f_\e$.}
\be 
\label{eq: wilsonian boundary action}
\PsiUV[\f_\e;\e]\equiv e^{-\SB[\f_\e;\e]}
\ee 
which satisfies the radial Schr\"odinger equation, $\partial_\e\PsiUV=-H_{RG}\PsiUV$ for the radial Hamiltonian $H_{RG}$ \cite{Mansfield:1999kk}; indeed, $\SB$ obeys the corresponding Hamilton--Jacobi equations in the classical limit \cite{Heemskerk:2010hk}. This radial-evolution form of the holographic RG flow is precisely why a connection with stochastic quantization was conjectured;\footnote{\label{footnote: SIR} If we stop viewing $S_{\text{bulk}}$ as an \emph{effective action} and treat it as UV-complete, then there is no inherent asymmetry between $z<\epsilon$ and $z>\epsilon$, and the full partition function can be written using the Heemskerk–Polchinski split \cite{Heemskerk:2010hk,Faulkner:2010jy} as
\be
\label{eq: HP split}
Z=\int\cD\f_\e\;\PsiUV[\f_\e;\e]\;\PsiIR[\f_\e;\e]
\ee
for $\PsiUV$ defined as before, and $\PsiIR$ defined analogously: integration of the bulk path integral for IR region $z>\e$ at fixed $\f_\e$. $\PsiIR$ satisfies an analogous (but ``time-flipped'') radial Schr\"odinger equation $\partial_\e\PsiIR=H_{RG}\PsiIR$, and $\SIR\equiv-\log\PsiIR$ satisfy the corresponding Hamilton--Jacobi equations.} indeed, one can now propose that \cite{Oh:2012bx,Jatkar:2013uga,Oh:2021bxx,Lee:2023ynb,Kim:2023bhp}
\be
\label{eq: dictionary}
\left(t,\;\SE,\;\psi_S\right)\quad\leftrightarrow\quad
\left(\e,\;2\SIR,\;\PsiUV\right)
\ee
where $\SIR$ is the on-shell action of the IR region with regular boundary conditions.\footnote{This is denoted as $-I_{os}$ in \cite{Oh:2012bx}; also see footnote~\ref{footnote: SIR} for more information on this object.}

This dictionary with the identification in \eqref{eq: definition of probability distribution} gives a relation between \emph{Wilsonian boundary action of AdS} and \emph{equilibrium Euclidean action} with \emph{probability distribution associated to its stochastic evolution in fictitious time}. This relation is best understood as a perturbative expansion around the Gaussian fixed point $\f=0$, which gives the schematic map
\be
\label{eq: multitrace dictionary}
D^{(n)}(\e)\evaluated_{\e=t}=P^{(n)}(t)-\half G^{(n)}\;.
\ee
for kernels $D^{(n)}$, $P^{(n)}$, and $G^{(n)}$ defined via\footnote{We follow the terminology of \cite{Oh:2012bx,Oh:2021bxx}: $D^{(n)}$ are the kernels of $\SB$ as a functional of the boundary value $\f_\e$ of the bulk field. The couplings of the multi-trace operators $\cO^n$ in the Wilsonian action of the boundary theory are obtained from them by the integral (at tree level, Legendre) transform with respect to $\f_\e$ \cite{Heemskerk:2010hk,Faulkner:2010jy}; for the double-trace coupling this gives a multiple of the inverse of $D^{(2)}$. We do not perform this transform.}
\be
\SB=\sum_n D^{(n)}\f^n,\; \SE=\sum_n G^{(n)}\f^n,\;
-\ln P=\sum_n P^{(n)}\f^n
\ee
where we are also stripping off a momentum-conserving delta function $\dhat(\sum_r\bm k_r)\equiv(2\pi)^d\de^d(\sum_r\bm k_r)$.

The map \eqref{eq: multitrace dictionary} has been checked for $n=2$ in free theories and for $n=3,4$ in the literature;\footnote{See footnote~\ref{footnote: cases that has been analyzed} for the relevant papers.} the main output of this paper is essentially the first explicit realization of this relation for an interacting gauge theory, including the gauge sector.

\subsection{Yang--Mills theory in \texorpdfstring{AdS$_4$}{AdS4}}
\label{sec: YM in AdS4}
The Yang--Mills action in Euclidean AdS$_4$ reads as
\begin{subequations}
\label{eq: YM action}
\begin{equation}
S=\frac14\int d^4x\sqrt{g}\,F^a_{\mu\nu}F^{a,\mu\nu},
\end{equation}
for $F^a_{\mu\nu}=\partial_\mu A^a_\nu-\partial_\nu A^a_\mu+g f^{abc}A^b_\mu A^c_\nu$, where we are explicitly working in the Poincar\'e patch $ds^2=(dz^2+d\bm x^2)/z^2$. Classical Weyl invariance rewrites this action as 
\begin{equation}
S=\frac14\int dz\,d^3x\,F^a_{\mu\nu}F^a_{\mu\nu}\;,
\end{equation}
with indices contracted with the flat metric: Yang--Mills theory in AdS$_4$ is classically equivalent to Yang--Mills theory on the flat half-space $z>0$, with the conformal boundary at $z=0$.
\end{subequations}

We will explicitly work in radial/axial gauge in this paper, i.e. $A^a_z=0$: the action then reads as
\be
\label{eq: YM radial gauge}
S=\int_0^\infty dz\int d^3x\left[\half\partial_zA^a_i\partial_zA^a_i+\half B^a_iB^a_i\right]
\ee
with $B^a_i=\half\e_{ijk}F^a_{jk}$. The radial ``time'' $z$ does not appear explicitly. The classical radial Hamiltonian, i.e.\ the Legendre transform of \eqref{eq: YM radial gauge}, is
\be
\label{eq: HRG}
\cH_{RG}=\int d^3x\left[\half E^a_iE^a_i-\half B^a_iB^a_i\right]\;,\qquad E^a_i=\partial_zA^a_i\;,
\ee
and the radial Schr\"odinger operator of \S~{\ref{sec: holographic wilsonian rg}} is $H_{RG}=\int d^3x\,[-\half\de^2/\de A^a_i\de A^a_i+\half B^a_iB^a_i]$. The residual gauge transformations are $z$-independent; since the Dirichlet condition requires the gauge parameter to vanish at $z=0$, only global colour rotations remain and the gauge is completely fixed \cite{Heemskerk:2010hk}. The $A_z$ equation of motion, i.e.\ the Gauss law $D_iE_i=0$, is not imposed in radial gauge; we return to this point in \S~{\ref{sec: gauge sector}}.

As commonly used in the relevant holography literature \cite{Raju:2011mp,Raju:2012zr,Raju:2012zs,Maldacena:2011nz,Bzowski:2013sza,Ghosh:2014kba,Anninos:2014lwa,Albayrak:2018tam,Albayrak:2019asr,Albayrak:2026auu}, we Fourier transform along the boundary, $A(z,\bm x)=\int_{\bm k} e^{i\bm k\.\bm x}A(z,\bm k)$ with $\int_{\bm k}\equiv\int d^3k/(2\pi)^3$, write $k\equiv\abs{\bm k}$, and introduce the transverse and longitudinal projectors
\be
\PiT_{ij}(\bm k)=\de_{ij}-\frac{k_ik_j}{k^2}\;,\qquad \PiL_{ij}(\bm k)=\frac{k_ik_j}{k^2}\;.
\ee
At quadratic order, $\half B^2$ becomes $\half A(-\bm k)\.k^2\PiT(\bm k)\.A(\bm k)$: transverse modes behave like conformally coupled scalars, while longitudinal modes are massless in the radial direction. With Dirichlet conditions at $z=0$ and regularity in the interior, the bulk-to-bulk propagator is \cite{Raju:2011mp,Albayrak:2018tam}
\be
\label{eq: gauge propagator}
G_{ij}(z,w;\bm k)=\PiT_{ij}(\bm k)\,\cG_{k}(z,w)+\PiL_{ij}(\bm k)\,\cG_0(z,w)\;,
\ee
where
\be
\label{eq: scalar propagator}
\cG_k(z,w)&=\frac{e^{-k|z-w|}-e^{-k(z+w)}}{2k}\;,\\
\cG_0(z,w)&=\min(z,w)\;,
\ee
are the propagator of a (rescaled) conformally coupled scalar and its $k\to0$ limit.

\section{Main tools}
\label{sec: main tools}
\subsection{Functional calculus of AdS$_4$ Yang--Mills}
\label{sec: Functional calculus}
The form of the bulk-to-bulk propagator in \eqref{eq: gauge propagator} and how it decomposes into scalar propagators via \eqref{eq: scalar propagator} motivate the following definition. For a function $W(k)$ of the norm of a momentum, let
\be
\label{eq: K operator}
\cK_{\bm k}[W]\equiv\PiT(\bm k)\,W(k)+\PiL(\bm k)\,W(0)=W(\Omega_{\bm k})\;,
\ee
with $\Omega_{\bm k}\equiv k\,\PiT(\bm k)$. In other words, $\cK$ is the functional calculus of the matrix $\Omega_{\bm k}$, whose eigenvalues are $k$ (transverse) and $0$ (longitudinal).\footnote{Essentially, this is simply a (for our purposes) more convenient way of writing the decomposition into straight and crossed lines of \cite{Albayrak:2019asr}.} In this language \eqref{eq: gauge propagator} reads $G=\cK_{\bm k}[\cG_k]$. Up to a factor of $-i$ and the Lorentzian boundary metric, $\cK$ is the operator
\be
\label{eq: projector}
\cP^{mn}_{\bm{k}}W\equiv-i\eta_{mn}W+i\frac{\bm{k}_m\bm{k}_n}{k^2}\left(W-W\big|_{\bm{k}\rightarrow0}\right)
\ee
of \cite{Albayrak:2019asr}. Since a functional calculus is an algebra homomorphism,
\be
\label{eq: homomorphism}
\cK_{\bm k}[W_1W_2]=\cK_{\bm k}[W_1]\.\cK_{\bm k}[W_2]\;,
\ee
which is the identity $\cP^{mn}_{\bm k}(AB)=i\cP^{m\a}_{\bm k}(A)\eta_{\a\b}\cP^{\b n}_{\bm k}(B)$ in Euclidean form, which follows from $\PiT\PiL=0$ and $(\PiT)^2=\PiT$, $(\PiL)^2=\PiL$. $\cK$ also commutes with radial integrations and sums. We will use \eqref{eq: homomorphism} repeatedly: any product of propagators or response functions carrying the same momentum is dressed by a single $\cK$.

\begin{figure}
\centering
$
\begin{aligned}
\includegraphics[scale=1.06]{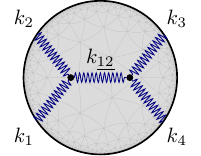}
\end{aligned}
\quad,\quad
\begin{aligned}
	\includegraphics[scale=.7]{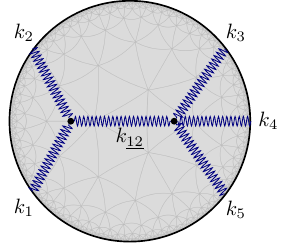}
\end{aligned}
$
\caption{\label{figure 1} Example Witten diagrams. Figure taken from \cite{Albayrak:2019asr}.}
\end{figure}

In radial gauge, the cubic and quartic terms in \eqref{eq: YM radial gauge} contain only boundary derivatives,
\be
\label{eq: YM interactions}
\half B^aB^a\supset{} g f^{abc}(\partial_iA^a_j)A^b_iA^c_j
+\frac{g^2}{4}f^{abe}f^{cde}A^a_iA^b_jA^c_iA^d_j\;,
\ee
so in momentum space the vertices are tensors $V^{abc}_{ijk}(\bm k_1,\bm k_2,\bm k_3)$ and $V^{abcd}_{ijkl}$ that do not depend on $z$; all radial dependence sits in the propagators \cite{Raju:2011mp}. This is why the expression for any gluon Witten diagram factorizes into vertex factors and a \emph{truncated diagram}, which depends on the external momenta only through the sum of their norms at each vertex \cite{Albayrak:2019asr}. For instance, the Witten diagrams in figure~{\ref{figure 1}} read
\be
\cW_4=V^{12m}V^{34n}\cM^{mn}_{\bm{k}_{{12}}}\;,\quad \cW_5=V^{12m}V^{345n}\cM^{mn}_{\bm{k}_{{12}}}\;,
\ee
where\footnote{
Here, $V^{12m}\equiv (\bm{\e}_1)_i(\bm{\e}_2)_jV^{ijm}$ for the polarization vectors $\bm{\e}_i$ (and similarly for others).} $\cM_{\bm{k}_{{12}}}$ is the truncated diagram with one bulk-to-bulk propagator of momentum $\bm{k}_{{12}}$.\footnote{We are using the convention of \cite{Albayrak:2018tam} to denote sums of momenta:
\begin{subequations}
\begin{multline}
k_{\underline{i_{11}i_{12}\dots i_{1n_1}}\;\underline{i_{21}i_{22}\dots i_{2n_2}}\dots \underline{i_{m1}i_{m2}\dots i_{mn_m}}j_1j_2\dots j_p}\\\coloneqq\sum\limits_{a=1}^{m}\abs{\sum\limits_{b=1}^{n_a}\bm{k}_{i_{ab}}}+\sum\limits_{c=1}^{p}\abs{\bm{k}_{j_c}}\;,
\end{multline}
and
\begin{equation}
\bm{k}_{i_1i_2\dots i_n}\coloneqq\bm{k}_{i_1}+\bm{k}_{i_2}+\cdots+\bm{k}_{i_n}\;.
\end{equation}
E.g., $k_{\underline{12}3\underline{45}}\equiv\abs{\bm{k}_1+\bm{k}_2}+\abs{\bm{k}_3}+\abs{\bm{k}_4+\bm{k}_5}\;$ and $\bm{k}_{12}\equiv\bm{k}_1+\bm{k}_2$.
\end{subequations}
} The crucial observation of \cite{Albayrak:2019asr} is that $\cM$ is the corresponding conformally coupled scalar diagram, dressed:
\be
\label{eq: dressed exchange}
\cM^{mn}=-i\,\cK^{mn}_{\bm{k}_{{12}}}\left[\frac{1}{k_{1234}\,k_{12\underline{12}}\,k_{34\underline{12}}}\right]\;,
\ee
where the bracket is the truncated exchange diagram of a conformally coupled scalar in AdS$_4$; more generally, every internal line is dressed by $\cK$.\footnote{In \cite{Albayrak:2019asr}, $W$ is computed with the diagrammatic method introduced in \cite{Arkani-Hamed:2017fdk} for dS wavefunctions, whereas similar computations were done by brute force in \cite{Albayrak:2018tam}. See also \cite{Oh:2020izq} for a compact result for conformally coupled scalars.} Since the factorization relies on Weyl invariance, it does not hold in AdS$_{d+1\neq4}$.

\subsection{Equilibrium actions of Yang--Mills in AdS$_4$}
\label{sec: equilibrium action}

\subsubsection{The regular branch}
\label{sec: regular branch}
Let $\SIR[A;\e]$ be the on-shell action of the region $z>\e$ with $A(\e)=A$ and regular behaviour as $z\to\infty$. By regular we mean that the action is finite, so that $A$ approaches a flat connection.\footnote{For transverse boundary data, this is the familiar decay $e^{-kz}$; longitudinal data approach a $z$-independent pure gauge.} Since the Lagrangian in \eqref{eq: YM radial gauge} has no explicit $z$-dependence, the shift $z\to z-\e$ shows that $\SIR$ does not depend on $\e$, i.e.\ $\partial_\e\SIR=0$. Its Hamilton--Jacobi equation, $\partial_\e\SIR=\cH_{RG}[A,E=-\de\SIR/\de A]$, then implies that the radial energy of a regular solution vanishes: with \eqref{eq: HRG}, $\SIR$ obeys the stationarity condition
\be
\label{eq: stationary HJ}
\half\int d^3x\left(\fdv{\SIR}{A^a_i}\right)^2=\half\int d^3x\, B^a_iB^a_i
\ee
for all $A$. Thus $\SIR$ is a zero-energy solution of the Hamilton--Jacobi equation, a ``superpotential'' in the sense of holographic RG flows \cite{deBoer:1999tgo,Freedman:2003ax,Skenderis:2006jq,Skenderis:2006rr}.

The regular solutions obey the first-order flow $\partial_zA=-\de\SIR/\de A$, which is the noiseless Langevin equation of \S~{\ref{sec: proposition}}. At tree level, $\SIR=-\log Z_{\rm CFT}[A]$ is the generating functional of connected gluon correlators, i.e.\ of the Witten diagrams of \cite{Albayrak:2019asr}. The equilibrium action of the dictionary \eqref{eq: dictionary} is therefore
\be
\label{eq: SE gluon}
\SE[A]=2\SIR[A]=-2\log Z_{\rm CFT}[A]\;,
\ee
and $e^{-\SE}=\PsiIR^2$ is the square of the regular radial wavefunction. Explicitly, to quartic order,
\bea[eq: SE kernels]
G^{(2)}&=\cK_{\bm k}[k]=\Omega_{\bm k}\;,\\
G^{(3)}&=\frac13\,V\circ\cK\left[\frac{1}{k_{123}}\right],\\
G^{(4)}&=-\frac14\Sym\left[VV\circ\cK\left[\frac{1}{k_{1234}k_{12\underline{12}}k_{34\underline{12}}}\right]\right]\nn\\
&\quad+\frac{1}{12}\,V^{(4)}\circ\cK\left[\frac{1}{k_{1234}}\right],
\eea
where $\cK$ dresses every line: each external leg $r$ by $\cK_r\equiv\cK_{\bm k_r}$ acting on the argument $k_r$, and, in the exchange term, the internal line by $\cK_{\underline{12}}$ acting on $k_{\underline{12}}$. The symbol $\circ$ contracts the vertex indices with the dressing operators of the corresponding lines, and $\Sym$ averages over permutations of the external labels (momentum, colour and vector index). At quadratic order $e^{-\SE}$ is Gaussian in the transverse modes and flat along the gauge orbits.

The kernels \eqref{eq: SE kernels} are finite whenever at least one external leg has a transverse component, which covers all the couplings computed in \S~{\ref{sec: flows}}.\footnote{If legs are purely longitudinal, the total energy $k_{1\cdots n}$ seen by \eqref{eq: SE kernels} vanishes as the profiles $\cK_{\bm k}[e^{-kz}]\,\bm\e=\PiL\bm\e$ of purely longitudinal legs do not decay. For $n=3$ this is harmless as $V$ vanishes on three longitudinal legs. For $n=4$ the individual diagrams are singular, but the singularities cancel as we can see by regulating the longitudinal profiles with $e^{-\de_rz}$: the internal longitudinal line then drops out, because its vertex numerator vanishes, and the residue of the $1/k_{1234}$ pole,
\be
\label{eq: longitudinal residue}
\mathrm{Res}=-\sum_{(ij)(kl)}\frac{V_{ij}\.\PiT(\bm q)\.V_{kl}}{\bm q^2}+V^{(4)}\;,
\ee
vanishes. Here the sum runs over the three pairings, $\bm q=\bm k_i+\bm k_j$, and $V_{ij}$ is the cubic vertex contracted with the polarizations of legs $i$ and $j$, with the internal colour summed over. This is the Ward identity of $\SE$.} Although not relevant for this paper, one does get divergent pieces in the middle steps of other computations; however, they do cancel among each other and lead to a finite final result. For instance, one such computation occurs in star-triangle topology of six-point gluon amplitude;\footnote{Please see the discussion in Appendix A of \cite{Albayrak:2019asr}.} a similar cancellation happens between the exchange and contact diagrams of purely-longitudinal four-point Witten diagram: schematically, the second-order magnetic field of a linearized pure gauge obeys $\bm q\.\bm B^{(2)}=0$, so the exchange term, which involves $\abs{\bm q\times\bm B^{(2)}}^2/\bm q^2$, cancels the contact term $\abs{\bm B^{(2)}}^2$.\footnote{Equivalently, the regular solution approaches a flat connection and the action density decays. The finite remainder,
\be
\label{eq: G4 longitudinal}
\bm\e_1\cdots\bm\e_4\.\de^4\SIR\big|_{\rm long}=\sum_{(ij)(kl)}\frac{V_{ij}\.\PiT(\bm q)\.V_{kl}}{\abs{\bm q}^3}\;,
\ee
is the action $\half\tilde A\.\abs{\bm q}\.\tilde A$ of the transverse field $\tilde A=\order{A^2}$ that completes a linearized pure gauge into a flat connection. It does not enter any of the transverse couplings below. We have verified these statements with $\SU(2)$ and $\SU(3)$ colours, in exact arithmetic, in the attached Mathematica file: the vanishing of $V$ on three longitudinal legs and of the longitudinal-line numerator, the cancellation \eqref{eq: longitudinal residue}, and the convergence of the regulated sum to \eqref{eq: G4 longitudinal}.}

\subsubsection{The Chern--Simons branch}
\label{sec: CS branch}
The stationarity condition \eqref{eq: stationary HJ} has solutions other than $\SIR$. Since $\de S_{CS}/\de A^a_i=B^a_i$ for the Chern--Simons functional
\be
\label{eq: CS functional}
S_{CS}[A]=\half\int d^3x\,\e_{ijk}\Big(A^a_i\partial_jA^a_k+\frac{g}{3}f^{abc}A^a_iA^b_jA^c_k\Big)\;,
\ee
$\pm S_{CS}$ solves \eqref{eq: stationary HJ} exactly. This is the familiar statement that the Chern--Simons state is a zero-energy solution of the Yang--Mills Schr\"odinger equation \cite{Witten:2003mb}. At quadratic order the stationary solutions are labelled by a sign for each helicity, $S^{(2)}=\half\sum_hs_h\,k\,\abs{A_h}^2$: $\SIR$ has $s_\pm=+1$, while $\pm S_{CS}$ has opposite signs for the two helicities. Holographically, this formal solution corresponds to boundary conditions that are regular in the IR for one helicity only.

Taking $\SE=\pm2S_{CS}$ in the dictionary of \eqref{eq: dictionary}, the drift becomes $-\half\de\SE/\de A_i=\mp B_i$. The noiseless Langevin flow is then the (anti-)self-duality equation $F_{ti}=\mp B_i$, i.e.\ instantons are the gradient flow of the Chern--Simons functional, and the supersymmetric (Parisi--Sourlas) form of the Fokker--Planck action \cite{Parisi:1979ka,Gozzi:1983qxk,Baulieu:1988jw} is Witten's topological Yang--Mills theory, viewed as supersymmetric quantum mechanics on the space of connections with the Chern--Simons functional as Morse function \cite{Witten:1988ze}. Since $e^{-\SE}$ is then not normalizable, it has no stochastic equilibrium. Moreover the linearized flow $\partial_tA=\mp\nabla\times A$ amplifies one helicity at all momenta, so the Langevin process is not well posed; the statements of this subsection are statements about the zero-energy state $e^{\mp S_{CS}}$ and about the noiseless flow. In the rest of the paper we take the regular branch, hence $\SE=2\SIR$.

\subsection{Tree-level correspondence to all orders}
\label{sec: proposition}

\subsubsection{Statement and its proof}

The relation \eqref{eq: multitrace dictionary} has been verified explicitly one multi-trace coupling at a time, and derived for general $n$, for scalar theories without explicit radial dependence, in section~4.1 of \cite{Oh:2021bxx} by completing the square in the bulk action. We restate that argument in a form that applies verbatim to gauge fields and makes its probabilistic content explicit.

\medskip\noindent\textbf{Proposition 1.}\footnote{This is the argument of \cite{Oh:2012bx,Oh:2021bxx}; see appendix~\ref{appendix: proposition} for what is and is not new.}~\emph{Consider a bulk theory whose action (for gauge fields in radial gauge) is $S=\int_0^\infty dz\int d^dx\,[\half(\partial_z\Phi)^2+\cV[\Phi]]$ with $\cV$ independent of $z$, and Dirichlet conditions $\Phi(0)=0$. Let $\SE=2\SIR$, and let $P[\Phi,t]$ be the distribution at time $t$ of the Langevin process $\partial_t\Phi=-\half\de\SE/\de\Phi+\eta$ with $\<\eta\eta\>=\de\de$ and $\Phi(0)=0$. Then, at tree level and up to a $\Phi$-independent constant $c$,
\be
\label{eq: proposition}
-\log P[\Phi,t]=\SB[\Phi;t]+\half\SE[\Phi]+c\;.
\ee
In particular, \eqref{eq: multitrace dictionary} holds for all $n$.}

\medskip\noindent\emph{Proof.} As in \eqref{eq: stationary HJ}, the $z$-independence of $\cV$ implies $\frac18\int(\de\SE/\de\Phi)^2=\int\cV[\Phi]$. The Onsager--Machlup representation of $P$ reads
\be
P[\Phi,t]=\int_{\Phi(0)=0}^{\Phi(t)=\Phi}\cD\Phi\;\exp\Big\{&-\half\int_0^tdt'\int d^dx
\\&\times\Big(\partial_{t'}\Phi+\half\fdv{\SE}{\Phi}\Big)^2\Big\}
\ee
which is same as \eqref{eq: fokker planck action} upto a term irrelevant at the tree level.\footnote{This is the last term in \eqref{eq: fokker planck action}, which comes from the Jacobian of the change of variables $\eta\to\phi$ (in the Stratonovich convention); it is of order $\hbar$ and does not contribute at tree level.} Expanding the square, the cross term integrates to $\half\SE[\Phi]-\half\SE[0]$, and the remaining terms give $\int_0^tdt'\int d^dx\,[\half(\partial_{t'}\Phi)^2+\cV[\Phi]]$, i.e.\ the bulk action of the region $0<z<t$. Hence $P\,e^{\SE/2}=\int_{\Phi(0)=0}^{\Phi(t)=\Phi}\cD\Phi\,e^{-S|_{0<z<t}}=e^{-\SB[\Phi;t]}$ at tree level. $\square$

\subsubsection{Interpretation}
\label{sec: interpretation}
Exponentiating \eqref{eq: proposition} and using \eqref{eq: wilsonian boundary action} and \eqref{eq: dictionary} gives $P[\Phi,t]=e^{-c}\,\PsiUV[\Phi;t]\,\PsiIR[\Phi]$. Normalizing $P$ fixes $e^{c}=\int\cD\Phi\,\PsiUV\PsiIR=Z$, the bulk partition function in the Heemskerk--Polchinski split \eqref{eq: HP split}. Hence
\be
\label{eq: P as HP integrand}
P[\Phi,t]=\frac{\PsiUV[\Phi;t]\,\PsiIR[\Phi]}{Z}
\ee
The Langevin process samples the bulk field on the slice $z=t$ with the Heemskerk--Polchinski weight. In other words, it is the ground-state (Doob) transform of the radial evolution by $\PsiIR$ \cite{Nelson:1966sp,Nelson:1973free,Glimm:1987ng,Damgaard:1987rr}: since $H_{RG}\PsiIR=0$, the drift is $\nabla\log\PsiIR=-\half\nabla\SE$. In Hamiltonian language, $\psi_S=Pe^{\SE/2}=\PsiUV/Z$, so $\psi_S$ and $\PsiUV$ obey the same equation with the same initial condition, i.e.\ $H_{FP}=H_{RG}$ at the classical level \cite{Oh:2012bx,Oh:2021bxx}.

At the non-perturbative level, the Doob-transform relation is an identity rather than a conjecture. If $\PsiIR$ is the exact positive ground state of $H_{RG}$, with energy $E_0$, and $\SE\equiv-2\log\PsiIR$, then $H_{FP}=H_{RG}-E_0$ holds identically, with $H_{FP}$ as in \eqref{eq: HFP}.\footnote{This is the standard ground-state transformation: $H_{RG}\PsiIR=E_0\PsiIR$ with $H_{RG}=-\half\nabla^2+\cV$ is $\cV=\frac18(\nabla\SE)^2-\frac14\nabla^2\SE+E_0$, which is \eqref{eq: HFP} up to $E_0$. With $\PsiIR$ taken $t$-independent, the constant in \eqref{eq: proposition} becomes $c=\log Z-E_0t$.} A loop-level dictionary is therefore a matter of defining $\PsiIR$ and $E_0$ (\S~{\ref{sec: outlook}}); what goes wrong if $\SE$ is instead kept at its tree-level value is the subject of remark~(d) below.

\subsubsection{Remarks}

\paragraph*{\textup{(a)} Scope.} The proposition applies to conformally coupled scalars in AdS$_{d+1}$ with the Weyl-invariant interaction $\f^{2(d+1)/(d-1)}$ after the rescaling $\f=z^{(d-1)/2}f$, and to Maxwell and Yang--Mills theory in AdS$_4$. The $\f^3$/AdS$_6$ and $\f^4$/AdS$_4$ checks of \cite{Oh:2021bxx,Lee:2023ynb} are instances. For fields not conformally coupled, $\SIR$ depends on $\e$; see remark~(c).

\paragraph*{\textup{(b)} Initial and boundary conditions.} Dirichlet conditions at $z=0$ correspond to $\Phi(0)=0$, i.e.\ $P[\Phi,0]=\de[\Phi]$. This is the setting of the Dirichlet analyses \cite{Oh:2021bxx,Lee:2023ynb}, and it implies that all multi-trace couplings with $n\geq3$ vanish at the boundary, $D^{(n)}(\e)\to0$ as $\e\to0$; see \S~{\ref{sec: flows}} for $n=3,4$.\footnote{We can heuristically understand this as follows. In the limit $\e\to0$ at fixed momenta, the tree-level solution in the slab $0<z<\e$ approaches $(z/\e)\Phi$. A term in the Lagrangian with $m$ $z$-derivatives and no explicit $z$-dependence then contributes $\order{\e^{1-m}}$ to $\SB[\Phi;\e]$. The kinetic term ($m=2$) gives the $1/\e$ divergence of $D^{(2)}$, while the radial-gauge interactions ($m=0$) give $\order{\e}$. In the Zwanziger gauges of \S~{\ref{sec: gauge sector}}, the cubic vertex $g f^{abc}X^a_iv^bA^c_i$ has $m=1$, but its leading $\order{1}$ part is symmetric in two legs and cancels against the antisymmetry of $f^{abc}$, so it is also $\order{\e}$. Hence, $D^{(n)}\to 0 $ for $n\ge 3$ in all these gauges.} Nothing in \eqref{eq: P as HP integrand} singles out this initial state. A general UV boundary condition $\PsiUV[\Phi;0]$, such as the Neumann and mixed conditions available for gauge fields in AdS$_4$ \cite{Witten:2003ya,Marolf:2006nd}, corresponds to the initial distribution $P[\Phi,0]\propto\PsiUV[\Phi;0]\,\PsiIR[\Phi]$, provided this product is normalizable; Dirichlet conditions are the special case $\PsiUV[\Phi;0]=\de[\Phi]$. Massive scalars with multi-trace boundary deformations were analyzed in \cite{Kim:2023bhp,Chae:2024zyn} by matching lower multi-trace couplings to the stochastic data at the initial time. Relating those matching conditions to this initial distribution is left to future work \cite{Albayrak:wip}.

\paragraph*{\textup{(c)} Radial inhomogeneity.} For bulk theories with explicit radial dependence, $\PsiIR$ depends on the cutoff. Then, by the Doob-transform argument of \S~{\ref{sec: interpretation}}, $P=\PsiUV\PsiIR/Z$ still solves a Fokker--Planck equation, now with the time-dependent drift $\nabla\log\PsiIR[\Phi;t]$.\footnote{This is the structure of the massive-scalar analysis of \cite{Chae:2024zyn}, whose Euclidean action has kernels that depend on the stochastic time; its two-point kernel is $-\partial_t\log$ of the regular radial solution. The earlier massive-scalar analyses \cite{Oh:2015xva,Kim:2023bhp} imposed additional conditions relating the boundary data to the initial stochastic time, and we have not attempted to relate them to the construction in this paper.} Nevertheless, $-\log P=\SB[\Phi;t]+\half\SE[\Phi;t]+c$ still holds, with $\SE[\Phi;t]\equiv2\SIR[\Phi;t]$, and so does $D^{(n)}=P^{(n)}-\half G^{(n)}(t)$. Therefore, radial homogeneity is not a condition for the existence of the dictionary; it is the condition under which the drift is time independent, so that the Langevin process is the stochastic quantization of a fixed action.

\paragraph*{\textup{(d)} Beyond tree level.} The identity $H_{FP}=H_{RG}-E_0$ of \S~{\ref{sec: interpretation}} needs $\SE$ to be the exact $-2\log\PsiIR$. Suppose instead that $\SE$ is kept at its tree-level value $\SE^{(0)}=2\SIR$. Stationarity gives $\frac18(\de\SE^{(0)}/\de\Phi)^2=\cV$ exactly, and hence
\be
\label{eq: HFP tree action}
H_{FP}=H_{RG}-\frac14\int d^dx\,\frac{\de^2\SE^{(0)}}{\de\Phi(x)\,\de\Phi(x)}\;.
\ee
For free fields the last term is a field-independent constant, and the dictionary holds to all orders. For interacting fields its field-dependent part acts as an extra one-loop potential, so the stochastic quantization of the tree-level action does not reproduce the one-loop holographic flow. For Yang--Mills theory the cubic term of $\SE^{(0)}$ does not contribute, since $f^{aac}=0$. The mismatch first appears in the double-trace coupling at $\order{g^2}$, through the trace of the quartic kernel $G^{(4)}$ over one pair of legs. The dictionary is restored by the one-loop correction $\SE^{(1)}$ of $-2\log\PsiIR$. Expanding $H_{FP}=H_{RG}-E_0$ shows that it obeys the transport equation\footnote{Equation \eqref{eq: transport} is the transport equation that accompanies the Hamilton--Jacobi equation \eqref{eq: stationary HJ} in the WKB expansion of the zero-energy Schr\"odinger equation, and it is solved by integrating along the regular classical solutions (\S~{\ref{app: transport}}): the one-loop $n$-point kernel is the tree $(n+2)$-point kernel with two legs sewn, divided by the total energy of the external legs. This is the Euclidean, Hamiltonian counterpart of the cutting rules for wavefunction coefficients, derived at tree level in \cite{Goodhew:2020ehn,Goodhew:2021oqg} and to all loop orders in \cite{Melville:2021lds} (see also the AdS Cutkosky rules of \cite{Meltzer:2020qbr} and the dS perturbative unitarity of \cite{Albayrak:2023hie}); see the Schr\"odinger-picture derivation of the one-loop rule in appendix~A of \cite{Melville:2021lds}, which builds on \cite{Cespedes:2020xqq}, with $E_0$ and the regularized traces in the role of the subtraction constants.}
\be
\label{eq: transport}
\int d^dx\,\fdv{\SE^{(0)}}{\Phi}\fdv{\SE^{(1)}}{\Phi}=\int d^dx\,\frac{\de^2\SE^{(0)}}{\de\Phi(x)\,\de\Phi(x)}-4E_0\;.
\ee
A one-loop comparison therefore tests the computation of $\SE^{(1)}$, i.e.\ of the one-loop ground state, rather than the dictionary itself. For Yang--Mills theory in AdS$_4$, the trace anomaly makes $\SE^{(1)}$ depend on the cutoff;\footnote{The Chern--Simons branch of \S~{\ref{sec: CS branch}} is the exception. There $\nabla^2S_{CS}=\int\de B^a_i/\de A^a_i=0$ identically, by $\e_{iji}=0$ and $f^{aac}=0$, so the last term in \eqref{eq: HFP tree action} vanishes and the hierarchy of \S~{\ref{app: transport}} terminates: $e^{\mp S_{CS}}$ is an exact zero-energy state \cite{Witten:2003mb}, $E_0=0$, and the ground-state transform of \S~{\ref{sec: interpretation}} with drift $\mp B_i$ holds to all orders, formally. It is the one branch on which the zero-energy condition holds exactly, without loop corrections; the corresponding Langevin flow, however, is not well posed (one helicity is amplified at all momenta), so this is a statement about the state $e^{\mp S_{CS}}$, not about a stochastic process.} however, the transport equation \eqref{eq: transport} actually makes this a tree-level calculation: we leave details of this computation and its consequences to future work \cite{Albayrak:wip}.

\section{The gauge sector}
\label{sec: gauge sector}
\subsection{Radial gauge is Parisi--Wu}
\label{sec: radial gauge}
We now combine the tools we have discussed in the previous section, \S~\ref{sec: main tools}. At quadratic order $G^{(2)}=\Omega_{\bm k}$, so the free Langevin equation is
\be
\label{eq: free langevin equation}
\partial_tA_i(t,\bm k)=-k\,\PiT_{ij}(\bm k)A_j(t,\bm k)+\eta_i(t,\bm k)\;,
\ee
with noise in all three components. Writing the solution with $A(0)=0$ as $A(t)=\int_0^tds\,R(t-s)\,\eta(s)$ defines the response function $R$, the retarded Green's function of \eqref{eq: free langevin equation}, $(\partial_\tau+\Omega_{\bm k})R=\de(\tau)$; the correlator is then $\int_0^{\min(t,t')}ds\,R(t-s)R(t'-s)^{T}$. By the functional calculus of \S~\ref{sec: Functional calculus}, the response function is
\begin{subequations}
\label{eq: free stochastic}
\be 
R_{ij}(\tau;\bm k)&=\cK_{\bm k}[e^{-k\tau}]_{ij}\,\theta(\tau)\;,
\ee 
for the Heaviside-theta distribution $\theta$, and the correlator reads
\be
\<A_i(t,\bm k)A_j(t',\bm k')\>&=\dhat(\bm k+\bm k')\,\cK_{\bm k}[\cG_k(t,t')]_{ij}\;,
\ee
\end{subequations}
i.e.\ $\PiT\cG_k(t,t')+\PiL\min(t,t')$. Transverse modes perform Ornstein--Uhlenbeck processes, while longitudinal modes perform Brownian motion with covariance $\min(t,t')=\cG_0(t,t')$.\footnote{Projected with $\PiT$ and $\PiL$, \eqref{eq: free langevin equation} reads $\partial_tA^\perp_i=-k\,A^\perp_i+\eta^\perp_i$ and $\partial_tA^\parallel_i=\eta^\parallel_i$. The first, a linear restoring drift with positive rate plus white noise, is the defining equation of the Ornstein--Uhlenbeck process, which relaxes to a Gaussian of variance $1/2k$; the second has no drift and is Brownian motion, whose variance $t$ grows without bound.} Comparing with \eqref{eq: gauge propagator}, the stochastic propagator coincides with the bulk-to-bulk propagator in radial gauge, including its longitudinal part. In other words, the term $W-W|_{\bm k\to0}$ of \eqref{eq: projector} is the random walk of the gauge modes in Parisi--Wu stochastic quantization. The same structure appears in the double-trace couplings. From the UV region one finds
\be
\label{eq: D2}
D^{(2)}(\e)=\half\cK_{\bm k}[k\coth k\e]=\half\Big[\PiT k\coth k\e+\frac{\PiL}{\e}\Big]\;,
\ee
while the Langevin process gives $P^{(2)}(t)=\half\<AA\>^{-1}=\cK_{\bm k}[k/(1-e^{-2kt})]$.\footnote{For a $U(1)$ field the longitudinal kernel $\PiL/(2\e)$ in \eqref{eq: D2} is the one found in \cite{Heemskerk:2010hk,Faulkner:2010jy}, where the corresponding mode is identified with the Goldstone (Wilson-line) mode of \cite{Nickel:2010pr}; in the present language, it is the Brownian motion of the gauge modes.} Hence $D^{(2)}=P^{(2)}-\half G^{(2)}$ holds separately in both sectors. The longitudinal coupling $1/(2\e)$ is half the inverse variance of a Brownian path, and it is absent from the gauge-invariant equilibrium distribution.

\subsection{The radial component of the gauge field is Zwanziger's gauge-fixing field}
\label{sec: gauge dictionary}
Now we relax the gauge condition. With $A_z\neq0$ the radial derivative is replaced by $F_{zi}=\partial_zA_i-D_iA_z$, and we consider the Langevin equation in the Baulieu--Zwanziger form \eqref{eq: BZ}, $F_{ti}=-\half\de\SE/\de A_i+\eta_i$, with $t=z$ and $A_t=A_z$. The Onsager--Machlup action becomes
\be
\half\int\Big(F_{ti}+\half\fdv{\SE}{A_i}\Big)^2&=\int\Big[\half F_{ti}F_{ti}+\frac18\Big(\fdv{\SE}{A_i}\Big)^2\Big]
\\&\quad+\half\int F_{ti}\fdv{\SE}{A_i}\;.
\ee
In the last term, $\int\partial_tA_i\,\de\SE/\de A_i=d\SE/dt$ as before, while $\int(D_iA_t)^a\,\de\SE/\de A^a_i=-\int A^a_t\,(D_i\,\de\SE/\de A_i)^a=0$ by gauge invariance of $\SE$. Hence the proof of Proposition~1 goes through for any $A_t$, and the Onsager--Machlup action is the bulk Yang--Mills action $\frac14\int F_{\mu\nu}F_{\mu\nu}$ with $A_z=A_t$. This gives the dictionary
\be
\label{eq: gauge dictionary}
A_z&\;\leftrightarrow\;v\;,\\
\text{bulk gauge condition}&\;\leftrightarrow\;\text{stochastic gauge fixing}\;,
\ee
under which bulk four-dimensional gauge invariance is the $(d+1)$-dimensional gauge invariance of \cite{Baulieu:1999wz}, radial gauge $A_z=0$ is Parisi--Wu, and the ``parabolic'' gauge $A_z=\a^{-1}\partial_iA_i$, with $\a>0$, is Zwanziger's gauge.\footnote{The stochastic side of this dictionary is the construction of \cite{Baulieu:1999wz}; what the present setting adds is that the equilibrium action is the non-local functional $\SE=2\SIR$ and that the extra dimension is the holographic one.} In the latter the longitudinal modes acquire a restoring drift and become Ornstein--Uhlenbeck processes with rate $k^2/\a$, so that the equilibrium distribution is normalizable; we work this out in \S~{\ref{sec: Zwanziger orders}}.

The dictionary has, formally, a third entry. If $A_z$ is integrated over rather than fixed, its equation of motion, the
Gauss law $D_iF_{zi}=0$, reads $D_i\eta_i=0$ on the Langevin side.\footnote{We do not attempt to define the corresponding constrained process; we only use the bulk statement, that integrating over $A_z$ imposes the Gauss law on the Wilsonian action, which we compare with the gauge-fixed prescriptions at quartic order in \S~{\ref{sec: gauge dependence}}. For an analysis of $U(1)$ fields on different choices for $A_z$ and of integrating over $A_z$, we refer the reader to \cite{Heemskerk:2010hk,Faulkner:2010jy}.} In radial gauge the Gauss law is not imposed, and correspondingly neither $\SB$ nor $P[A,t]$ is gauge invariant at finite $t$; only gauge-invariant data equilibrate. With $A_z$ integrated over, the holographic Wilsonian action is gauge invariant, and we compare the two prescriptions at quartic order in \S~{\ref{sec: gauge dependence}}. This extends to the stochastic side, and to the interacting non-abelian theory, the treatment of the gauge sector of \cite{Heemskerk:2010hk,Faulkner:2010jy}, which was left out of the $U(1)$ analysis of \cite{Oh:2012bx}.

\subsection{Faddeev--Popov determinant and Gribov copies}
\label{sec: FP}
The classical argument of \S~{\ref{sec: gauge dictionary}} survives in the path integral because the gauge-fixing measure is trivial, as in the five-dimensional formulation of \cite{Baulieu:1999wz}, whose arguments we adapt to the slab with a Dirichlet condition. In the bulk path integral, the parabolic gauge comes with the Faddeev--Popov operator $M_{FP}=D_z-\a^{-1}\partial_iD_i\equiv\partial_z+K$, with $K=g\,\mathrm{ad}(A_z)-\a^{-1}\partial_iD_i$. It acts on gauge parameters with $\w(0)=0$, which are the transformations that preserve the Dirichlet condition. These are not required to vanish at $z=\e$, so they act on the cutoff data, consistent with the fact that $\SB$ is not gauge invariant (\S~{\ref{sec: gauge dependence}}). Since $M_{FP}$ is first order in $z$ with an initial condition, it is a Volterra operator, and $\det M_{FP}=\exp\big(\theta(0)\int dz\,\mathrm{Tr}\,K\big)$. Moreover, $\mathrm{Tr}\,K$ is field independent, because $\mathrm{Tr}\,\mathrm{ad}(X)=f^{aac}X^c=0$ for both $X=A_z$ and the $g\,\partial_i\mathrm{ad}(A_i)$ part of $\partial_iD_i$. The determinant is therefore a (formally divergent) constant, and no ghosts are needed, in agreement with stochastic gauge fixing, where no determinant arises in the first place \cite{Zwanziger:1981kg,Baulieu:1981ec}.

In perturbation theory, the gauge condition is reached by solving a forward parabolic equation for $\w$ from $\w(0)=0$; this is well posed for $\a>0$. Its solution is unique, order by order in $g$, so there are no Gribov copies \cite{Zwanziger:1981kg,Baulieu:1999wz}. As in radial gauge, the condition together with $\w(0)=0$ fixes $\w$ completely, in particular $\w(\e)$, so there is no residual gauge freedom. Beyond perturbation theory, the question is existence rather than uniqueness: smooth solutions of the nonlinear flow for finite gauge transformations are unique whenever they exist, but the flow must exist for all $z$ for every configuration to be brought to the gauge. For the group-valued gauge transformation $U(z,\bm x)$ its principal part, $U^{-1}\partial_zU=\a^{-1}\partial_i(U^{-1}\partial_iU)$, is the harmonic-map heat flow into the gauge group, which can develop finite-time singularities in three dimensions \cite{ChenDing:1990}. We do not address this question, which plays no role at tree level. The Langevin Jacobian is likewise field independent up to the term $-\frac14\de^2\SE/\de A^2$ of \eqref{eq: HFP}, which matters only at loop level (\S~{\ref{sec: outlook}}).

\subsection{Zwanziger's gauge at quadratic and cubic order}
\label{sec: Zwanziger orders}
At quadratic order the drift in Zwanziger's gauge is $\Omega^{(\a)}_{\bm k}=k\PiT+\a^{-1}k^2\PiL$, and the dressing operator becomes the functional calculus of $\Omega^{(\a)}$,
\be
\label{eq: K alpha}
\cK^{(\a)}_{\bm k}[W]=\PiT(\bm k)\,W(k)+\PiL(\bm k)\,W(k^2/\a)\;.
\ee
This interpolates between the radial-gauge operator of \cite{Albayrak:2019asr} ($\a\to\infty$) and a gauge in which the longitudinal modes are frozen ($\a\to0$). The equilibrium distribution is $e^{-\SE-\a^{-1}\int(\partial_iA_i)^2}$ at quadratic order. In the bulk, the longitudinal part of $\half F_{zi}F_{zi}$ becomes $\half\abs{(\partial_z+\g)A^\parallel}^2$ with $\g=k^2/\a$. Its cross term is a total derivative, which contributes a boundary term to $\SB$:
\be
\label{eq: D2 alpha}
D^{(2)}_\a(\e)&=\half\cK^{(\a)}_{\bm k}[k\coth k\e]+\frac{k^2}{2\a}\PiL\;,\\
P^{(2)}_\a(t)&=\cK^{(\a)}_{\bm k}\Big[\frac{k}{1-e^{-2kt}}\Big]\;.
\ee
These again satisfy $D^{(2)}_\a=P^{(2)}_\a-\half G^{(2)}$, with the gauge-invariant $G^{(2)}=\Omega_{\bm k}$. The longitudinal double-trace coupling now flows from the Brownian value $1/(2\e)$ in the UV to the gauge-fixing term $k^2/\a$ in the IR, exponentially fast: the Zwanziger damping removes the power-law tails of radial gauge, please see \eqref{eq: F0 power law} and \S~{\ref{sec: gauge dependence}}.

At cubic order, for non-abelian fields, both sides acquire an extra term. On the Langevin side it is the quadratic part $g f^{abc}A^b_iv^c$ of the Zwanziger drift $D_iv$. In the bulk it is the vertex $g f^{abc}X^a_iv^bA^c_i$ in $\half F_{zi}F_{zi}$, with $X_i=\partial_zA_i-\partial_iv$ the linear part of $F_{zi}$. Neither term is a gradient, but the Onsager--Machlup argument of \S~{\ref{sec: gauge dictionary}} does not require one anyway. We can proceed by defining $\mathscr{D}_3\equiv\bm\e_1\bm\e_2\bm\e_3\.\de^3\SB/\de A_1\de A_2\de A_3$, and similarly $\mathscr{P}_3$ and $\mathscr{G}_3$. With the profiles $\bm u_r(z)=\cK^{(\a)}_{\bm k_r}[S_{k_r}(z)]\,\bm\e_r$, see \eqref{eq: UV propagators}, the holographic side gives
\be
\label{eq: D3 alpha}
\mathscr{D}^{(\a)}_3(\e)={}&\int_0^\e dz\,\Big[V(\bm u_1,\bm u_2,\bm u_3)
\\&-g\sum_{\s\in S_3}f^{a_{\s1}a_{\s2}a_{\s3}}\,(\bm X_{\s1}\.\bm u_{\s2})\,v_{\s3}\Big]\;,
\ee
where $V(\bm u_1,\bm u_2,\bm u_3)$ is the cubic vertex of \eqref{eq: YM interactions} evaluated on the profiles, $\bm X_r=(\partial_z+\frac{k_r^2}{\a}\PiL_r)\bm u_r$ and $v_r=\a^{-1}i\bm k_r\.\bm u_r$. The Langevin side gives
\be
\label{eq: P3 alpha}
\mathscr{P}^{(\a)}_3(t)=\sum_{r=1}^3\int_0^tds\,\Big[\half\mathscr{G}_3(\bm x_r,\bm u_{r'},\bm u_{r''})-Z_r\Big]\;,
\ee
where $\bm x_r=(\partial_s+\Omega^{(\a)}_{\bm k_r})\bm u_r$ is the amputated response function, $\{r,r',r''\}=\{1,2,3\}$, and $Z_r=g f^{a_ra_{r'}a_{r''}}(\bm x_r\.\bm u_{r'})\,v_{r''}+(r'\leftrightarrow r'')$ comes from the Zwanziger drift.\footnote{By the argument of \S~{\ref{sec: gauge dictionary}}, \eqref{eq: D3 alpha} and \eqref{eq: P3 alpha} must satisfy $\mathscr{D}^{(\a)}_3=\mathscr{P}^{(\a)}_3-\half\mathscr{G}_3$. We have verified this in exact rational arithmetic for various $\a$ values, with generic polarizations, with one or two purely longitudinal legs, and on transverse data; dropping either cubic term violates it at order one, as detailed in the ancillary Mathematica file.} For transverse external data $v_r=0$, and $\mathscr{D}^{(\a)}_3$ coincides with \eqref{eq: D3} for every $\a$, whereas the couplings with longitudinal legs depend on $\a$. This does not extend to higher orders: transverse linear data are not gauge invariant beyond linear order, and at quartic order the internal line of the exchange diagrams can be longitudinal. The quadruple-trace couplings on transverse data then depend on $\a$ through a single zero mode (\S~{\ref{sec: gauge dependence}}).

\section{Multi-trace flows of gluons}
\label{sec: flows}
\subsection{Reduction to scalar seeds}
Proposition~1 already implies \eqref{eq: multitrace dictionary} for gluons at tree level. It is nevertheless instructive, and useful for explicit results, to see how the gluon couplings arise from scalar ones. On the holographic side, the UV region $0<z<\e$ has the bulk-to-cutoff propagator $\cK_{\bm k}[S_k(z)]$ and the bulk-to-bulk propagator $\cK_{\bm k}[\cG^{\rm UV}_k]$, where
\be
\label{eq: UV propagators}
S_k(z)&=\frac{\sinh kz}{\sinh k\e}\;,\\
\cG^{\rm UV}_k(z_1,z_2)&=\frac{\sinh(kz_<)\sinh(k(\e-z_>))}{k\sinh k\e}\;.
\ee
for $z_<\coloneqq\min(z_1,z_2)$ and $z_>\coloneqq\max(z_1,z_2)$. The longitudinal limit $\cG^{\rm UV}_0=\min(z,w)-zw/\e=z_<(\e-z_>)/\e$ is the covariance of a Brownian bridge.\footnote{Brownian bridge is the Brownian motion of the gauge modes in \S~{\ref{sec: radial gauge}}, conditioned to return to zero at $z=\e$; likewise, $\cG^{\rm UV}_k$ is the covariance of an Ornstein--Uhlenbeck bridge.} On the stochastic side, tree diagrams are built from the response function and correlator \eqref{eq: free stochastic} and from the vertices of $\SE$. On both sides the vertices are $z$- (or $t$-)independent tensors, so by \eqref{eq: homomorphism} every tree diagram equals the vertex tensors contracted with $\cK$-dressed radial integrals of a scalar theory with the same diagram topology. The relevant scalar seeds are the cubic contact and exchange diagrams, whose radial integrals coincide with those of the conformally coupled $\f^3$ theory in AdS$_6$ \cite{Oh:2021bxx}, and the quartic contact diagram of the conformally coupled $\f^4$ theory in AdS$_4$ \cite{Lee:2023ynb}. Since the scalar relations hold as identities in the norms of all momenta, including their $k\to0$ limits, they lift to gluons leg by leg.

\subsection{Triple-trace couplings}
To first order in $g$, the UV region gives
\be
\label{eq: D3}
D^{(3)}(\e)&=\frac{1}{3!}\,V\circ\cK_1\cK_2\cK_3\big[I_3(\e)\big]\;,\\
I_3(\e)&=\int_0^\e dz\prod_{r=1}^3S_{k_r}(z)\;,
\ee
with the closed form\footnote{The equation \eqref{eq: I3} is known in the literature: it (along with its quadruple-trace sibling $I_4$) are the seeds of \cite{Oh:2021bxx,Lee:2023ynb}. On the other hand, the exchange seed $F(\e)$ and its longitudinal limit with the power-law tail that we discuss below (see \eqref{eq: F closed}, \eqref{eq: F0 power law}, and \eqref{eq: J}) were absent in the literature.}
\be
\label{eq: I3}
I_3(\e)=\frac{\sum\limits_{\eta\in\{\pm1\}^2}\eta_2\eta_3\,(\cosh\s_\eta\e-1)/\s_\eta}{4\sinh k_1\e\,\sinh k_2\e\,\sinh k_3\e}\Bigg|_{\substack{\s_\eta= k_1+\eta_2k_2+\eta_3k_3}}\;,
\ee
As $\e\to\infty$,\footnote{We note that the classical scale invariance of the kernels imply $D^{(n)}=\e^{n-3}f_n(\bm k_i\e)$, so the infrared limit $\e\to\infty$ at fixed momenta is also the limit of large momenta in units of the cutoff.} $I_3\to1/k_{123}$, so that $D^{(3)}\to\half G^{(3)}$. On the stochastic side, the connected three-point function at first order in $g$ gives\footnote{This follows from
\be 
\label{eq: P3 extraction}
P^{(3)}=-\<AAA\>_c/(3!\prod\<AA\>)\;.
\ee 
}
\be
\label{eq: P3}
P^{(3)}(t)=\frac{1}{3!}\,V\circ\cK_1\cK_2\cK_3\Big[I_3(t)+\frac{1}{k_{123}}\Big]\;.
\ee
This follows from the scalar identity
\be
\label{eq: total derivative identity}
\sum_{r=1}^3\int_0^tds\,\frac{R_{k_r}(t-s)}{g_{k_r}(t)}\prod_{r'\neq r}\frac{\cG_{k_{r'}}(s,t)}{g_{k_{r'}}(t)}=1+k_{123}\,I_3(t)\;,
\ee
where $R_k(\tau)=e^{-k\tau}$ and $g_k(t)=\cG_k(t,t)$. The identity holds because $R_k(t-s)/g_k(t)=(\partial_s+k)S_k(s)$ and $\cG_k(s,t)/g_k(t)=S_k(s)$ (with $\e\to t$ in $S_k$), so that the left-hand side integrates to a total derivative.\footnote{The same argument gives, for any number $n$ of legs, $\sum_{r=1}^n\int_0^tds\,\frac{R_{k_r}(t-s)}{g_{k_r}(t)}\prod_{r'\neq r}S_{k_{r'}}(s)=1+k_{1\cdots n}I_n(t)$ with $I_n(t)=\int_0^tds\prod_rS_{k_r}(s)$, which is the contact identity used at quartic order below.} Thus we arrive at $D^{(3)}=P^{(3)}-\half G^{(3)}$, consistent with Proposition~1. Note that $P^{(3)}$ is the cubic on-shell action of the full half-space with the field pinned at $z=t$, in line with \S~\ref{sec: interpretation}.\footnote{Explicitly, see equation \eqref{eq: P as HP integrand}.}

\subsection{Quadruple-trace couplings: seeds}
At second order the UV region gives exchange and contact terms,
\be
\label{eq: D4}
D^{(4)}(\e)=-\frac18\Sym\Big[VV\circ\cK\big[F(\e)\big]\Big]
+\frac{1}{4!}V^{(4)}\circ\cK\big[I_4(\e)\big]\;,
\ee
with
\be 
I_4(\e)=\int_0^\e dz\prod_{r=1}^4S_{k_r}(z)
\ee 
and
\be
\label{eq: F}
F(\e)=\int_0^\e\!\!\int_0^\e dzdwS_{k_1}(z)S_{k_2}(z)
\cG^{\rm UV}_{k_{\underline{12}}}(z,w)S_{k_3}(w)S_{k_4}(w).
\ee
The stochastic quadruple-trace weight is extracted from the connected four-point function via
\be
\label{eq: P4 extraction}
P^{(4)}=\frac{1}{4!}\Big[\sum_{\rm channels}(3!)^2P^{(3)}\<AA\>P^{(3)}-\frac{\<AAAA\>_c}{\prod\<AA\>}\Big]\;.
\ee
At order $g^2$ the four-point function receives three kinds of contributions: diagrams with a response function as internal line (from iterating the cubic drift), diagrams with a correlator as internal line, and the quartic term of $\SE$. For the scalar seeds we have verified that they combine into
\be
\label{eq: P4}
P^{(4)}(t)=&-\frac18\Sym\Big[VV\circ\cK\Big[F(t)+\frac{1}{k_{1234}k_{12\underline{12}}k_{34\underline{12}}}\Big]\Big]
\\&+\frac{1}{4!}V^{(4)}\circ\cK\Big[I_4(t)+\frac{1}{k_{1234}}\Big]\;,
\ee
i.e.\ $D^{(4)}=P^{(4)}-\half G^{(4)}$ via \eqref{eq: SE kernels}. The contact part follows from the four-leg version of \eqref{eq: total derivative identity}; this is the identity behind \cite{Lee:2023ynb}. The exchange part is guaranteed by Proposition~1;\footnote{ We have also checked it at rational values of the momenta, where both sides are computed exactly from the Langevin equation and from the slab and agree to the 60 digits displayed, for generic momenta and in the limits where the internal line or an external leg becomes longitudinal (see the attached Mathematica file).} at the fixed point, the $\order{\l^2}$ exchange contributions of a bulk cubic coupling were already considered in \cite{Oh:2020zvm,Oh:2021bxx}. As far as we are aware, for an exchange generated by a bulk cubic vertex, this comparison has not been made at finite cutoff before (see \cite{Kim:2023bhp,Chae:2024zyn} for exchange-type contributions of multi-trace deformations).

We can now proceed to obtaining closed form expressions as both seeds are elementary. Expanding $\prod_{r}\sinh k_rz$ into hyperbolic cosines gives
\be
\label{eq: I4}
I_4(\e)=\frac{\sum\limits_{\eta\in\{\pm1\}^3}\eta_2\eta_3\eta_4\,\sinh(\s_\eta\e)/\s_\eta}{8\sinh k_1\e\,\sinh k_2\e\,\sinh k_3\e\,\sinh k_4\e}\Bigg|_{\substack{\s_\eta= k_1+\eta_2k_2\\\quad+\eta_3k_3+\eta_4k_4}}\;,
\ee
For the exchange seed, with $s\equiv k_{\underline{12}}$,
\be
\label{eq: F closed}
F(\e)=\frac{\sum_{\eta_2,\eta_4=\pm1}\eta_2\eta_4\;\cJ_s(k_1+\eta_2k_2,\,k_3+\eta_4k_4)}{4\sinh k_1\e\,\sinh k_2\e\,\sinh k_3\e\,\sinh k_4\e}\;,
\ee
whereas the longitudinal seed $F^{(0)}$ follows from $F^{(0)}\equiv F|_{s\to0}$. Here $\cJ_s(a,b)$ is an elementary function: see \S~{\ref{app: seeds}} for its details, along with those of the longitudinal seed $F^{(0)}$.

In the IR, we have $I_4\to1/k_{1234}$ and $F\to1/(k_{1234}k_{12\underline{12}}k_{34\underline{12}})$ exponentially fast, whereas the longitudinal seed approaches its limit only as a power,
\be
\label{eq: F0 power law}
F^{(0)}(\e)=\frac{1}{k_{1234}k_{12}k_{34}}-\frac{1}{\e\,k_{12}^2k_{34}^2}+\dots\;,
\ee
where the dots are exponentially small. We derive \eqref{eq: F0 power law} in \S~{\ref{sec: gauge dependence}}.\footnote{The $1/\e$ term is the zero mode of the Brownian bridge $\cG^{\rm UV}_0$, i.e.\ it comes from the diffusion of the gauge modes in radial gauge, just as the longitudinal double-trace coupling $\PiL/(2\e)$ in \eqref{eq: D2}. It is therefore a property of radial gauge rather than of the gauge-invariant sector. } 

In the UV, $I_4=\e/5+\order{\e^3}$ and $F,F^{(0)}=\e^3/112+\order{\e^5}$, so all quadruple-trace couplings vanish at the boundary, as required by the Dirichlet condition, and the contact term dominates near it.

\subsection{Colour-ordered quadruple-trace flows}
\label{sec: colour ordered flows}

\subsubsection{Expressions}

We can now contract quadruple-trace couplings with transverse polarizations, $\bm\e_r\.\bm k_r=0$, and define the quadruple-trace vertex $\mathscr{D}_4\equiv\bm\e_1\bm\e_2\bm\e_3\bm\e_4\.\de^4\SB/\de A_1\cdots\de A_4=4!\,D^{(4)}\.\bm\e_1\cdots\bm\e_4$. In the DDM basis of \S~{\ref{app: color}},
\be
\mathscr{D}_4={}&\mathscr{D}_4(1,2,3,4;\e)\,f^{a_1a_2c}f^{ca_3a_4}
\\&+\mathscr{D}_4(1,3,2,4;\e)\,f^{a_1a_3c}f^{ca_2a_4}\;,
\ee
with the colour-ordered quadruple-trace flows
\be
\label{eq: colour ordered flows}
\mathscr{D}_4(1,2,3,4;\e)&=g^2\big[N_s(\e)-N_t(\e)+C_{1234}\,I_4(\e)\big]\;,\\
\mathscr{D}_4(1,3,2,4;\e)&=g^2\big[N_t(\e)-N_u(\e)+C_{1324}\,I_4(\e)\big]\;.
\ee
The $s$-channel function is
\be
\label{eq: channel functions}
N_s(\e)&=n_s^\perp F_s(\e)+n_s^\parallel F^{(0)}_s(\e)\;,\\
n_s^\perp&=v_{12}\.\PiT(\bm k_1+\bm k_2)\.v_{34}\;,\\
n_s^\parallel&=-\frac{(\bm\e_1\.\bm\e_2)(\bm\e_3\.\bm\e_4)(k_1^2-k_2^2)(k_3^2-k_4^2)}{\abs{\bm k_1+\bm k_2}^2}\;,
\ee
with $F_s$ the exchange seed for the pairing $(12)(34)$ and
\be
\label{eq: v12}
v_{12}=(\bm\e_1\.\bm\e_2)(\bm k_1-\bm k_2)+2(\bm\e_1\.\bm k_2)\,\bm\e_2-2(\bm\e_2\.\bm k_1)\,\bm\e_1\;,
\ee
and similarly for $v_{34}$. The $t$- and $u$-channel functions follow by successive $(243)$ relabellings, i.e.\ the pairings $(14)(23)$ and $(13)(42)$. The contact polynomials are
\be
\label{eq: contact polynomials}
C_{1234}={}&2(\bm\e_1\.\bm\e_3)(\bm\e_2\.\bm\e_4)-(\bm\e_1\.\bm\e_2)(\bm\e_3\.\bm\e_4)
\\&-(\bm\e_1\.\bm\e_4)(\bm\e_2\.\bm\e_3)\;,
\ee
and $C_{1324}$ is obtained by exchanging $2\leftrightarrow3$. The longitudinal part $n^\parallel$ is the AdS-specific Coulomb-like term: it involves $k_r^2=\abs{\bm k_r}^2$ and would vanish for null external momenta. In the IR it has only the total-energy pole, $n_s^\parallel/(k_{1234}k_{12}k_{34})=-(\bm\e_1\.\bm\e_2)(\bm\e_3\.\bm\e_4)(k_1-k_2)(k_3-k_4)/(\abs{\bm k_1+\bm k_2}^2k_{1234})$, as befits an instantaneous interaction (\S~{\ref{sec: gauge dependence}}). 

\subsubsection{Comments}

We can interpret the results in the light of Proposition~1:
\be 
\mathscr{P}_4(\cdots;t)=\mathscr{D}_4(\cdots;t)+\half\mathscr{G}_4=\mathscr{D}_4(\cdots;t)+\mathscr{D}_4(\cdots;\infty)
\ee 
where the second equality uses that the slab couplings approach the equilibrium kernel in the infrared, \eqref{eq: D4 IR}. As for $P^{(3)}$, this is the quartic on-shell action of the half-space with the field pinned at $z=t$: the slab $0<z<t$ contributes $\mathscr{D}_4(t)$ and the region $z>t$ contributes $\half\mathscr{G}_4$, see \eqref{eq: P as HP integrand}.

We can also comment on the infrared and flat-space limits.\footnote{Besides being consistency checks, these limits are important from a different perspective as well: the finite-cutoff couplings on transverse data are not gauge invariant; transverse data themselves are not gauge invariant beyond linear order. Their gauge-independent content is their infrared limit, i.e.\ the gluon correlators and their flat-space limit, and the way the dependence on the bulk gauge condition is confined to a single zero mode as we discuss further in \S~{\ref{sec: gauge dependence}}.} In radial gauge, as $\e\to\infty$,
\be
\label{eq: D4 IR}
\mathscr{D}_4(1,2,3,4;\infty)={}&g^2\Big[\frac{n_s^\perp}{k_{1234}k_{12\underline{12}}k_{34\underline{12}}}+\frac{n_s^\parallel}{k_{1234}k_{12}k_{34}}
\\&-(s\to t)+\frac{C_{1234}}{k_{1234}}\Big]\;,
\ee
which is the colour-ordered combination $W_s-W_t$ of the gluon Witten diagrams \eqref{eq: dressed exchange} of \cite{Albayrak:2019asr} plus the contact diagram, up to the overall sign of $\de^4\SIR=-\<JJJJ\>_c$ and the factor of $-i$ in their conventions. As an independent check of all relative signs, the residue at the total-energy pole $k_{1234}\to0$ must be the flat-space colour-ordered amplitude for the null momenta $p_r=(\bm k_r,i\abs{\bm k_r})$ \cite{Maldacena:2011nz,Raju:2012zr,Raju:2012zs}. Indeed, at $k_{1234}=0$ the transverse and longitudinal parts combine into the spatial part of the flat-space axial-gauge propagator, $\PiT/(k_{12\underline{12}}k_{34\underline{12}})+\PiL/(k_{12}k_{34})\to(\mathbb{1}-\bm q\bm q/k_{12}^2)/(\bm q^2-k_{12}^2)$ with $\bm q=\bm k_1+\bm k_2$. At random complex kinematics with $k_{1234}=0$ and helicity polarizations, we numerically verified to high precision that the residue vanishes for the all-plus and single-minus configurations, as it must for tree-level Yang--Mills amplitudes. For MHV configurations, the ratio of the two colour orderings reproduces the Parke--Taylor ratio $-\<13\>\<24\>/(\<12\>\<34\>)$ \cite{Parke:1986gb}.

\subsection{Gauge dependence and the Gauss law}
\label{sec: gauge dependence}

The longitudinal channel of the quadruple-trace flows isolates the gauge dependence, and we can understand it in four steps:
\begin{enumerate}
\item \emph{In radial gauge}, the longitudinal exchange is the Coulomb energy of the radial colour charges minus a zero mode of the Brownian bridge; the Coulomb term is instantaneous and carries the total-energy pole, the zero mode is what relaxes as a power.
\item \emph{In Zwanziger's gauge}, the Coulomb term is unchanged and only the zero mode changes: it relaxes exponentially, but to an $\a$-dependent infrared value, so the equilibrium distribution of Zwanziger's process is not the naive gauge-fixed weight beyond quadratic order, while the flat-space limit is unaffected. 
\item \emph{The same infrared shift} follows from the stationary Fokker--Planck equation of Zwanziger's process alone, which provides a non-trivial consistency check of our quartic formulae for a drift that is not a gradient. 
\item \emph{Finally}, the zero mode is the remnant of the constraint that the gauge condition imposes on $E^\parallel$; it is absent when the Gauss law is imposed, and it is a redundant coupling that a gauge rotation of the cutoff data removes. The content of the flows independent of the bulk gauge condition is therefore the transverse exchange, the contact term and the Coulomb term.
\end{enumerate}

\subsubsection{Coulomb term and zero mode}
In radial gauge the longitudinal exchange splits into a local Coulomb term and a zero mode. For the pairing $(12)(34)$, let
\be
\label{eq: Wronskian}
W_{12}(z)\equiv \partial_zS_{k_1}(z)\,S_{k_2}(z)-S_{k_1}(z)\,\partial_zS_{k_2}(z)\;.
\ee
Up to normalization, $(\bm\e_1\.\bm\e_2)W_{12}$ is the colour-charge density $gf^{abc}A^b_i\partial_zA^c_i$ of the two transverse modes, i.e.\ the source of the Gauss law $D_iE_i=0$.\footnote{Explicitly, $\displaystyle W_{12}(z)=\frac{(k_1-k_2)\sinh(k_{12}z)-k_{12}\sinh((k_1-k_2)z)}{2\sinh k_1\e\,\sinh k_2\e}$.} Since $\partial_z^2S_k=k^2S_k$, we have $\partial_zW_{12}=(k_1^2-k_2^2)S_{k_1}S_{k_2}$. The longitudinal projection of the cubic vertex is therefore $(\bm\e_1\.\bm\e_2)\partial_zW_{12}/\abs{\bm k_1+\bm k_2}$, and $n_s^\parallel$ in \eqref{eq: channel functions} is the product of two such factors. Integrating by parts against the Brownian bridge and using \eqref{eq: bridge identity}, we find
\be
\label{eq: Coulomb zero mode radial}
n_s^\parallel F_s^{(0)}={}&-\frac{(\bm\e_1\.\bm\e_2)(\bm\e_3\.\bm\e_4)}{\abs{\bm k_1+\bm k_2}^2}\bigg[\int_0^\e\!dz\,W_{12}W_{34}\\
&-\frac1\e\int_0^\e\!dz\,W_{12}\int_0^\e\!dw\,W_{34}\bigg]\;.
\ee
The first term is the Coulomb energy of the radial colour charges. It is local in $z$, i.e.\ instantaneous in stochastic time, and it carries the total-energy pole. The second term is the $-zw/\e$ part of the bridge covariance, its zero mode. The internal longitudinal field vanishes at both ends of the slab, so in radial gauge $E^\parallel=\partial_zA^\parallel$ has zero mean, $\int_0^\e E^\parallel=0$, and cannot follow the total charge $\int_0^\e W_{12}$. As $\e\to\infty$ the two integrals approach $(k_1-k_2)(k_3-k_4)/k_{1234}$ and $(k_1-k_2)/k_{12}$ exponentially fast, so the power law \eqref{eq: F0 power law} comes entirely from the zero mode.

\subsubsection{Zwanziger's gauge}
In Zwanziger's gauge the Coulomb term is unchanged and the entire gauge dependence sits in the zero mode. With $A_z=\a^{-1}\partial_iA_i$ two things change. The internal longitudinal line becomes the Ornstein--Uhlenbeck bridge $\cG^{\rm UV}_\g$ with $\g\equiv\abs{\bm k_1+\bm k_2}^2/\a$, see \eqref{eq: K alpha}. In addition, the vertex $gf^{abc}X^a_iv^bA^c_i$ of \S~{\ref{sec: Zwanziger orders}} couples to this line through its $v$, which turns the longitudinal source into $(\partial_z-\g)W_{12}$. By using \eqref{eq: bridge identity}, we can see that the longitudinal channel function becomes
\begin{multline}
\label{eq: Coulomb zero mode alpha}
N^{\parallel}_{s,\a}={}-\frac{(\bm\e_1\.\bm\e_2)(\bm\e_3\.\bm\e_4)}{\abs{\bm k_1+\bm k_2}^2}\bigg[\int_0^\e\!dz\,W_{12}W_{34}\\
-\frac{\g}{\sinh\g\e}\,\hat W_{12}\hat W_{34}\bigg]\;,
\end{multline}
with $\hat W_{12}\equiv\int_0^\e dz\,e^{\g(z-\e/2)}W_{12}(z)$. The transverse internal line and the contact term are unchanged, since the Zwanziger vertices need a longitudinal leg. Hence $\mathscr{D}^{(\a)}_4(1,2,3,4;\e)$ is given by \eqref{eq: colour ordered flows}, with $n_s^\parallel F^{(0)}_s$ replaced by $N^\parallel_{s,\a}$ and similarly in the $t$ channel. As $\a\to\infty$ the zero-mode term reduces to the radial one. For finite $\a$ it approaches its limit exponentially fast, so Zwanziger's gauge has no power-law tails. Its infrared value, however, differs from that of radial gauge:
\be
\label{eq: D4 alpha IR}
\mathscr{D}^{(\a)}_4(1,2,3,4;\infty)=\mathscr{D}_4(1,2,3,4;\infty)+g^2\big(Z_s-Z_t\big)\;,
\ee
where
\be
\label{eq: Z alpha}
Z_s=\frac{2(\bm\e_1\.\bm\e_2)(\bm\e_3\.\bm\e_4)}{\a}\,\frac{k_1-k_2}{k_{12}+\g_s}\,\frac{k_3-k_4}{k_{34}+\g_s}\;,
\ee
with $\g_s=\abs{\bm k_1+\bm k_2}^2/\a$, and $Z_t$ is obtained by the relabelling used in \eqref{eq: channel functions}. $Z_s$ has no total-energy pole, so the flat-space limit of \S~{\ref{sec: colour ordered flows}} does not depend on $\a$. By \eqref{eq: multitrace dictionary} the same shift appears in the equilibrium weights, $\mathscr{P}^{(\a)}_4(1,2,3,4;\infty)=2\mathscr{D}_4(1,2,3,4;\infty)+g^2(Z_s-Z_t)$, where the first term is the colour-ordered part of the gauge-invariant kernel $\mathscr{G}_4$. The equilibrium distribution of Zwanziger's process is therefore not $e^{-\SE-\a^{-1}\int(\partial_iA_i)^2}$ beyond quadratic order:\footnote{This is expected as the gauge-fixing force is not conservative \cite{Baulieu:1999wz}; \eqref{eq: Z alpha} above gives the first correction explicitly.} transverse data see the nonlinear completion of the gauge-fixing weight through the internal longitudinal line.\footnote{We have verified equations \eqref{eq: Coulomb zero mode radial}--\eqref{eq: Z alpha} against a brute-force evaluation of the exchange diagrams with $\SU(2)$ colours, both cubic vertices and the Zwanziger-gauge propagator: see the attached Mathematica file for details.}

\subsubsection{The shift from the stationary Fokker--Planck equation}
The shift $Z_s-Z_t$ can be derived on the stochastic side alone, without reference to the bulk. We first write the equilibrium distribution of Zwanziger's process as $e^{-\SE-\Delta}$ and use the fact that gauge invariance implies $\int(D_iv)^a\,\de\SE/\de A^a_i=0$; then, at tree level, $\Delta$ obeys
\be
\label{eq: stationary FP}
\Big(\half\frac{\de \SE}{\de A}+Dv\Big)\.\frac{\de \Delta}{\de A}+\half\abs{\frac{\de \Delta}{\de A}}^2=0\;,
\ee
with the dot denoting the integral over $\bm x$ and the sums over colour and vector indices. At quadratic order $\Delta_2=\a^{-1}\int(\partial_iA_i)^2$. At higher orders, the linear part of $\half\,\de\SE/\de A+Dv$ together with $\de\Delta_2/\de A$ multiplies each monomial by the sum of the relaxation rates of its legs, $k$ for a transverse and $\g$ for a longitudinal one, and the equation is solved order by order.\footnote{On transverse data, $\Delta_3$ has one longitudinal leg. Its term with legs $1$ and $2$ is $-2\g_s\,\bm j_{12}\.\bm A^\parallel/(k_{12}+\g_s)$, where $\bm j_{12}$ is the longitudinal projection of $\half\de\SE^{(3)}/\de A$ on legs $1$ and $2$; the Zwanziger term $(Dv)_2$ does not contribute, since $v$ vanishes on transverse legs. At quartic order the cubic part of the drift multiplies $\de\Delta_2/\de A$, which vanishes on transverse data, and the four legs have the total rate $k_{1234}$. The terms $(\half\,\de\SE^{(3)}/\de A)\.\de\Delta_3/\de A$ and $\half\abs{\de\Delta_3/\de A}^2$ combine into $k_{1234}$ times a separable expression, which gives \eqref{eq: FP Delta4}.} On transverse data the result at quartic order is
\be
\label{eq: FP Delta4}
\bm\e_1\cdots\bm\e_4\.\de^4\Delta=\sum_{(ab)(cd)}\frac{2\g_{ab}\;\bm j_{ab}\.\bm j_{cd}}{(k_{ab}+\g_{ab})(k_{cd}+\g_{ab})}\;,
\ee
with $\g_{ab}=\abs{\bm k_a+\bm k_b}^2/\a$ and $\bm j_{ab}$ the longitudinal projection of $\half\de\SE^{(3)}/\de A$ on legs $a$ and $b$. This reproduces $g^2(Z_s-Z_t)$ for the colour ordering $(1,2,3,4)$, as we have checked with $\SU(2)$ colours in the Mathematica file. The cancellation of $k_{1234}$ is the stochastic counterpart of the absence of a total-energy pole. Only the cubic kernel of $\SE$ enters, and the agreement with \eqref{eq: D4 alpha IR}--\eqref{eq: Z alpha} checks the gauge-sector dictionary of \S~{\ref{sec: gauge sector}} at quartic order, for a drift that is not a gradient.

\subsubsection{Gauss law and redundancy}
The zero mode depends on the bulk gauge condition: it vanishes when the Gauss law is imposed, and it can be removed by a gauge rotation of the cutoff data. As $\a\to0$, the gauge condition becomes the Coulomb gauge $\partial_iA_i=0$, with $A_z$ fixed by the Gauss law, and the zero-mode term vanishes: $Z_s=\order{\a}$, and so is the zero-mode term in \eqref{eq: Coulomb zero mode alpha} at finite $\e$. The same result follows if $A_z$ is integrated over instead of fixed. The longitudinal field $E^\parallel=\partial_zA^\parallel-\partial^\parallel A_z$ is then unconstrained, and minimizing $\half\int\abs{E^\parallel}^2$ against the charge source leaves only the Coulomb term. The zero-mode terms in \eqref{eq: Coulomb zero mode radial} and \eqref{eq: Coulomb zero mode alpha} are what remains of the constraints $\int_0^\e E^\parallel=0$ and $\int_0^\e e^{\g z}E^\parallel=0$. These are imposed by the two gauge conditions, under which $E^\parallel=\partial_zA^\parallel$ and $E^\parallel=(\partial_z+\g)A^\parallel$ respectively. With the Gauss law imposed, the holographic Wilsonian action is independent of the bulk gauge condition. Moreover, minimizing the Wilsonian action of any bulk gauge over gauge rotations $V$ of the cutoff data gives the Gauss-law action, $S^{\rm G}_{\rm B}[A]=\min_VS_{\rm B}[A^V]$. 

This follows from a simple observation: every slab configuration with $A_z$ unconstrained is a gauge transformation, with $U(0)=1$, of a configuration in the chosen gauge whose cutoff data are $A^V$ for some $V$. The zero-mode couplings can therefore be removed by the redefinition $A\to A^{V_*[A]}$ of the cutoff data, with $V_*$ the minimizing rotation. This redefinition is non-local, and so is the resulting Gauss-law action: already at quadratic order only the sum $\PiT k\coth k\e+\PiL/\e$ of \eqref{eq: D2} is analytic at $\bm k=0$. This is the non-abelian counterpart of the observation of \cite{Heemskerk:2010hk,Faulkner:2010jy} that the constrained amplitude is not a local Wilsonian action and that the longitudinal (Wilson-line) mode has to be retained in it; the radial-gauge kernels, including the zero mode and its power-law relaxation, are the ones with the expected locality. The quadruple-trace couplings of the Gauss-law action on transverse data relax exponentially, and their infrared limit is that of radial gauge. In this sense the part of the quadruple-trace flows that does not depend on the bulk gauge condition is the transverse exchange, the contact term and the Coulomb term, while the power-law relaxation in radial gauge reflects the unrestricted random walk of the gauge modes in Parisi--Wu stochastic quantization.

\section{The de Sitter reading}
\label{sec: dS reading}

Although the dictionary of this paper is Euclidean, it has a Lorentzian counterpart in de Sitter space and there is a well established and widely studied analytical continuation between them \cite{Maldacena:2002vr,Harlow:2011ke,Sleight:2021plv,DiPietro:2021sjt}. It is then natural to expect that our results can actually be useful in the context of cosmology, and we review that perspective in this section.

\subsection{Wavefunction, in-in correlators and stochastic inflation}
\label{sec: dS structure}
Euclidean AdS$_4$ in the Poincar\'e patch continues to the expanding patch of dS$_4$ by $z\to-i\eta$ together with the continuation of the radius, and for Weyl-invariant fields both are the flat half-space in conformal time. Under this continuation the regular solution $e^{-kz}$ becomes the Bunch--Davies mode \cite{Bunch:1978yq}, so the regular radial wavefunction $\PsiIR[\Phi;\e]$ becomes the Bunch--Davies wavefunction $\Psi_{\rm BD}[\Phi;\eta]$, and $\SIR$ becomes the generating functional of its coefficients $\psi_n$. The equilibrium weight of the Langevin process,
\be
e^{-\SE}=\PsiIR^2\quad\leftrightarrow\quad\abs{\Psi_{\rm BD}}^2\;,
\ee
is the Euclidean form of the Born-rule weight whose moments are the equal-time in-in correlators. The extraction formulas \eqref{eq: P3 extraction} and \eqref{eq: P4 extraction}, which convert the kernels of $-\log P$ into connected correlators, are the Euclidean form of the familiar relations between wavefunction coefficients and in-in correlators \cite{Maldacena:2002vr,Weinberg:2005vy,Cespedes:2020xqq,Palma:2026qgn}, $\<\phi\phi\>=1/(2\,\mathrm{Re}\,\psi_2)$, $\<\phi\phi\phi\>_c=-2\,\mathrm{Re}\,\psi_3/\prod(2\,\mathrm{Re}\,\psi_2)$, and so on; here all quantities are real and the real parts are absent. Equilibrium correlators of the Langevin process are thus the Euclidean counterparts of in-in correlators.

The finite-time distribution $P=\PsiUV\PsiIR/Z$ of \eqref{eq: P as HP integrand} continues to the product of $\Psi_{\rm BD}[\Phi;\eta]$ with the amplitude to evolve from the configuration $\Phi$ at time $\eta$ to a vanishing field on the future boundary $\eta=0$, which is what the Dirichlet condition at $z=0$ becomes. This is a bridge, not a quantity usually computed in cosmology, and the finite-cutoff couplings $D^{(n)}(\e)$ are its kernels: they describe the relaxation of the process from the sharp initial state $\de[\Phi]$ to the in-in equilibrium, not the wavefunction coefficients at finite time, which are $\e$-independent for Weyl-invariant fields. The in-in weight itself corresponds to the equilibrium initial condition of remark~(b), $\PsiUV[\Phi;0]\propto\PsiIR[\Phi]$, for which $P=e^{-\SE}$ at all $t$. A Lorentzian construction that prepares the bra from the future, by extending the Poincar\'e patch, has recently been given in \cite{Donath:2024utn}; the Heemskerk--Polchinski split is its Euclidean analogue, with the slab in the role of the future preparation.

The ground-state transform of \S~{\ref{sec: interpretation}} also has a Lorentzian counterpart. For a real-time wavefunction $\Psi(\eta)$, Nelson's stochastic mechanics \cite{Nelson:1966sp} constructs a real diffusion whose density is $\abs{\Psi(\eta)}^2$, with drift $\nabla\log\abs{\Psi}$ plus the current velocity $\nabla\arg\Psi$; the Doob transform used here is the special case of a real wavefunction, for which the current vanishes. The relation between Nelson's diffusion for the Bunch--Davies wavefunction and the Langevin equation of stochastic inflation, whose noise comes from modes crossing the horizon, will be discussed in \cite{Albayrak:wip}.

Stochastic inflation is where this structure is already in use. For a light scalar in de Sitter space, the long-wavelength modes perform a Langevin process in cosmic time, and their distribution obeys a Fokker--Planck equation whose equilibrium distribution is $P_{\rm eq}\propto e^{-8\pi^2V/3H^4}$ \cite{Starobinsky:1986fx,Starobinsky:1994bd}, the exponential of the Euclidean action on the four-sphere \cite{Mirbabayi:2019qtx}. In recent years the Fokker--Planck equation and its corrections have been derived from quantum field theory in de Sitter space: from the Bunch--Davies wavefunction, by integrating out the short modes of $P=\Psi^*_{\rm BD}\Psi_{\rm BD}$ \cite{Gorbenko:2019rza}; in the static patch, by tracing out the environment of a subhorizon observer \cite{Mirbabayi:2019qtx,Mirbabayi:2020vyt}; and as the dynamical renormalization group of Soft de Sitter Effective Theory \cite{Cohen:2020php,Cohen:2021fzf}.

The identification of the holographic Wilsonian flow with a Langevin process is the Euclidean, anti-de Sitter version of the statement above, with two differences. In de Sitter space the Fokker--Planck equation is an effective description of the long-wavelength modes, obtained by splitting the field at a comoving cutoff, integrating out the short modes and approximating the result; the real-time evolution itself is oscillatory. In Euclidean signature the radial evolution $e^{-zH_{RG}}$ is a heat kernel, so the radial Schr\"odinger equation is a diffusion equation for all modes and at every $z$, and $P=\PsiUV\PsiIR/Z$ obeys the Fokker--Planck equation exactly, with nothing integrated out. The second difference concerns the time variable. The continuation $z\to-i\eta$ produces conformal time, in which the Bunch--Davies wavefunction of a light scalar depends on $\eta$, so that in the variables of this paper its drift is time dependent, the regime of remark~(c); stochastic inflation obtains a fixed drift and a fixed equilibrium action by a different route, using the number of e-folds as time and keeping only super-horizon modes. Weyl invariance makes the drift time independent in conformal time for all modes, which is Proposition~1, and it also removes secular infrared growth; for conformally coupled scalars and gluons in dS$_4$ the stochastic approach therefore has no infrared content. Its content for gluons is the gauge sector, to which we now turn.

\subsection{Gluons in dS$_4$}
\label{sec: dS gluons}
Yang--Mills theory in dS$_4$ is flat-space Yang--Mills theory in conformal time, and its late-time wavefunction coefficients have been computed in the same axial-gauge momentum-space formalism as the anti-de Sitter correlators of \cite{Albayrak:2019asr}; see \cite{Albayrak:2026auu} and references therein. Up to the continuation factors, the equilibrium kernels \eqref{eq: SE kernels} are twice these wavefunction coefficients, since $e^{-\SE}=\PsiIR^2$. The equilibrium correlators of the Langevin process are then the Euclidean counterparts of the in-in gluon correlators. The colour-ordered infrared couplings \eqref{eq: D4 IR}, their total-energy pole and the flat-space amplitude at the pole are the familiar statements of the cosmological bootstrap \cite{Arkani-Hamed:2018kmz,Baumann:2019oyu,Baumann:2020dch,Baumann:2022jpr} in this language.

The gauge-sector results translate as follows. Temporal gauge, $A_\eta=0$, which is the standard choice in cosmological gluon computations, is radial gauge, and by \S~{\ref{sec: radial gauge}} it is Parisi--Wu stochastic quantization: the longitudinal propagator, $\min(z,z')$ up to the continuation factors, of these computations is the covariance of the Brownian motion of the gauge modes, which do not equilibrate. Gauge-invariant equilibrium data are unaffected by this, as in Parisi--Wu, but the finite-time distribution is not gauge invariant, and the entire gauge dependence of the quartic couplings on transverse data sits in the longitudinal zero mode of \S~{\ref{sec: gauge dependence}}. The continuation of Zwanziger's gauge, $A_\eta=(i\a)^{-1}\partial_iA_i$, damps the gauge modes of the Euclidean process without ghosts and without Gribov copies in perturbation theory (\S~{\ref{sec: FP}}); the price is that the equilibrium distribution on transverse data differs from the naive gauge-fixed one by the shift $Z_s-Z_t$ of \eqref{eq: Z alpha}, which reflects the fact that transverse modes are not gauge invariant beyond linear order. Imposing the Gauss law, or minimizing over gauge rotations of the data, removes the zero mode and leaves the gauge-invariant content of \S~{\ref{sec: gauge dependence}}. We expect the same bookkeeping to be needed in a stochastic treatment of the gauge potentials, as opposed to the physical field strengths used in \cite{Talebian:2019opf,Fujita:2022fit}, in particular for the non-abelian gauge fields of the inflationary models reviewed in \cite{Maleknejad:2012fw} (gauge-flation, chromo-natural inflation), whose fluctuations propagate on a time-dependent homogeneous background and therefore belong to the time-dependent-drift regime of remark~(c).

Finally, the Chern--Simons branch of \S~{\ref{sec: CS branch}} is parity odd, with opposite signs for the two helicities, and its gravitational cousin is the Kodama state \cite{Kodama:1990sc} discussed alongside the Chern--Simons state in \cite{Witten:2003mb}; we comment more on this in the next section.

\section{Outlook}
\label{sec: outlook}

\emph{De Sitter space.} The Lorentzian version of the dictionary is partly in place and partly not. The part that is in place is \S~{\ref{sec: dS structure}}: $P=\abs{\Psi_{\rm BD}}^2$ is the Born rule, and for light scalars the Fokker--Planck description of its long-wavelength modes, with its corrections, has been derived from quantum field theory in de Sitter space \cite{Gorbenko:2019rza,Mirbabayi:2019qtx,Mirbabayi:2020vyt,Cohen:2020php,Cohen:2021fzf}. What this paper adds on the Euclidean side, the all-order tree-level proof for radially homogeneous bulk theories and the treatment of the gauge sector, has no counterpart there in this form. Although stochastic formalisms for gauge fields during inflation exist \cite{Prokopec:2007ak,Talebian:2019opf,Fujita:2022fit}, they are written for the long-wavelength physical electric and magnetic fields of abelian models: a treatment along the lines of this paper would in addition have to handle  the random walk of the gauge modes in temporal gauge, the Gauss law, and a damping gauge of Zwanziger's type, which is what \S~{\ref{sec: gauge sector}} and \S~{\ref{sec: gauge dependence}} do in the Euclidean setting, and it would have to do so with a time-dependent drift: it would be interesting to see this correspondence made quantitative.

\emph{Loops.} Proposition~1 is a tree-level statement, and we have not attempted a loop analysis in this paper. Such an analysis would have to address three points.
\begin{enumerate}
\item[(i)] It would have to define the exact $\PsiIR$, i.e.\ the operator ordering and measure of the bulk radial path integral, including the Faddeev--Popov sector of the chosen bulk gauge. Given $\PsiIR$, the Jacobian term $-\frac14\de^2\SE/\de A^2$ of \eqref{eq: HFP}, i.e.\ the It\^o--Stratonovich choice, is fixed by the identity $H_{FP}=H_{RG}-E_0$ of \S~{\ref{sec: interpretation}}.
\item[(ii)] It would have to regularize the ultraviolet-divergent functional traces, such as $\mathrm{Tr}\,\Omega$ and $\mathrm{Tr}\,\de^2\SE/\de A^2$, which determine $E_0$, compatibly on both sides.
\item[(iii)] Beyond tree level, Yang--Mills theory in AdS$_4$ is in any case not equivalent to its flat-space counterpart: its phase structure depends on the coupling and on the boundary conditions \cite{Aharony:2012jf}. It would have to account for the one-loop running of the Yang--Mills coupling. As the trace anomaly breaks the Weyl invariance that made $\SIR$ independent of $\e$, this would mean that $\PsiIR$ would acquire a logarithmic dependence on the cutoff, hence one should consider Langevin processes beyond time-homogeneous domain.
\end{enumerate}

\emph{Gravitons.} Analysis of time-inhomogeneous Langevin processes is already a prerequisite even at tree level for the analysis of gravitons, as Einstein gravity in AdS$_4$ is not Weyl invariant with the graviton bulk-to-boundary propagator in axial gauge being $(1+kz)e^{-kz}$ \cite{Albayrak:2019yve,Albayrak:2020bso}. Therefore, a stochastic description of graviton multi-trace flows requires the time-dependent construction of remark~(c) together with a treatment of the gravitational constraints analogous to the Gauss law. Since graviton Witten diagrams in AdS$_4$ have been computed in the same momentum-space, axial-gauge formalism \cite{Albayrak:2019yve,Albayrak:2020bso}, we expect explicit tree-level checks of the graviton analogue of \eqref{eq: multitrace dictionary} to be actually feasible: we hope to address this in the future.

\emph{Chern--Simons branch.} It would be interesting to relate the Chern--Simons branch of \S~{\ref{sec: CS branch}} to the self-dual limit of Yang--Mills theory in AdS$_4$ studied in \cite{Heuveline:2026knw}. There the $\theta$-angle enters through the boundary conditions \cite{Witten:2003ya,Marolf:2006nd}; in our language, it leads to complex weights and complex Langevin dynamics \cite{Parisi:1983mgm}. On the de Sitter side, the branch has the Kodama state as its gravitational cousin \cite{Witten:2003mb}, which ties it to the de Sitter and graviton items above: it would be interesting to see such progress at the intersection of cosmology, gravity, and stochastic quantization.

\begin{acknowledgments}
The author would like to thank Savan Kharel for insightful discussions. Claude LLM Opus 5.5 made partial contributions at late stages of this work.\footnote{Essentially in aiding the author with the verification of the main results, in preparation and polishing of the attached \texttt{Mathematica} file, and further general technical and editorial assistance.} The author
takes full responsibility for the final result. In the early stages of this research, the author was supported by a VIDI grant of the Netherlands Organisation for Scientific Research (NWO) that is funded by the Dutch Ministry of Education, Culture and Science (OCW). In the later stages, he was supported by the BAGEP Award of the Science Academy in Türkiye and by the TUBITAK (The Scientific and Technological Research Council of
Türkiye)\footnote{Please see ROR ID: \href{https://ror.org/04w9kkr77}{https://ror.org/04w9kkr77}} 2232-b International Fellowship for Early Stage Researchers program, project number 122c153.
\end{acknowledgments}

\appendix

\section{Appendices}

\subsection{The all-order correspondence in earlier work}
\label{appendix: proposition}

Proposition~1 builds on several earlier works, and we want to be precise about what it adds.

\subsubsection{Previous literature}
Mansi, Mauri and Petkou conjectured that the bulk action is the Fokker--Planck action of a stochastic process in radial time, with the classical boundary action equal to twice the boundary effective action \cite{Mansi:2009mz}. They observed that this can hold only for special actions. They verified the identification for the conformally coupled $\f^4$ theory in AdS$_4$, using its exact bulk solutions and invoking the Stokes-theorem structure of stochastic quantization \cite{Baulieu:1988jw}.

Oh and Jatkar recast the conjecture as an equality between the Fokker--Planck and radial Hamiltonians, with $\psi_S=Pe^{\SE/2}$ identified with $\PsiUV=e^{-\SB}$ and $\SE=-2I_{os}$, and described it as partially proved by their examples \cite{Oh:2012bx}.

The general argument is due to Oh \cite{Oh:2021bxx}: for a scalar bulk action whose Lagrangian has no explicit radial dependence (the conformally coupled scalar in AdS$_6$ being the worked example), the on-shell boundary action obeys the zero-energy Hamilton--Jacobi equation, the bulk action is a perfect square up to a total derivative, it coincides with the Fokker--Planck action in the classical limit once $t=r$ and $\SE=2I_{os}$, and $\SB=-\ln P-\half\SE$, i.e.\ \eqref{eq: multitrace dictionary}, follows for all $n$. Together with the Onsager--Machlup representation of $P$ with a sharp initial state given in \cite{Oh:2012bx}, this is the content of the proof of Proposition~1. The explicit checks in \cite{Oh:2021bxx} were made for the three-point function.

Further examples have since been checked \cite{Jatkar:2013uga,Lee:2023ynb,Kim:2023bhp,Chae:2024zyn}. For massive scalars, the Euclidean action of \cite{Chae:2024zyn} has kernels that depend on the stochastic time.

\subsubsection{Main contribution}
We do not provide a new relation with Proposition~1. What we add is (i) its formulation as a statement about wavefunctions, $P=\PsiUV\PsiIR/Z$, which identifies the Langevin process as the ground-state (Doob) transform of the radial evolution (\S~{\ref{sec: interpretation}}) --- the classic correspondence between Euclidean field theory and ground-state diffusions \cite{Nelson:1966sp,Nelson:1973free,Glimm:1987ng}, which appears in stochastic quantization as $e^{-S}=\abs{\Psi_0}^2$ \cite{Dijkgraaf:2009gr}; (ii) the consequences of this formulation: the initial distribution that corresponds to a general ultraviolet boundary condition (remark~(b)), the fact that without radial homogeneity the same construction holds with a time-dependent drift (remark~(c), as in \cite{Chae:2024zyn}), and the form of the loop-level statement (remark~(d)); and (iii) its extension to gauge fields, where the gauge-fixed drift is not a gradient (\S~{\ref{sec: gauge sector}}).

Radial homogeneity gives the stationarity $\partial_\e\SIR=0$ of the radial Hamilton--Jacobi equation. Completing the square in the Onsager--Machlup action then reproduces the bulk action, and the Stokes term accounts for the shift $-\half G^{(n)}$. This completion is the Bogomol'nyi, or ``fake superpotential'', structure of holographic RG flows \cite{deBoer:1999tgo,Freedman:2003ax,Skenderis:2006jq,Skenderis:2006rr}. The proof also identifies the Langevin process as the ground-state (Doob) transform of the radial evolution (\S~{\ref{sec: interpretation}}). This is the classic correspondence between Euclidean field theory and ground-state diffusions \cite{Nelson:1966sp,Nelson:1973free,Glimm:1987ng}, which appears in stochastic quantization as $e^{-S}=\abs{\Psi_0}^2$ \cite{Dijkgraaf:2009gr}. Two consequences follow. First, stationarity is not needed for the existence of the dictionary: without it, the same construction holds with a drift that depends on the stochastic time, as in \cite{Chae:2024zyn} (remark~(c)). What stationarity ensures is that $\SE$ is a fixed action, so that the Langevin process is a stochastic quantization in the usual sense. Second, the construction fixes the initial data (remark~(b)), and it extends to gauge fields whose gauge-fixed drift is not a gradient (\S~{\ref{sec: gauge sector}}).

\subsection{Colour decomposition in Yang--Mills}
\label{app: color}
Let us consider a gauge theory with three-point and four-point interactions. For the antisymmetric structure constants $f^{abc}$, any $n$-point Witten diagram reads as
\be
\label{eq: DDM decomposition}
\cW_{\a}^{a_1\dots a_n}={}&W_{\a}\sum\limits_{\pi\in S_{n-2}}b_{\a,\pi}\, f^{a_1a_{\pi(2)}c_1}f^{c_1a_{\pi(3)}c_2}
\\&\times\cdots f^{c_{n-3}a_{\pi(n-1)}a_n}\;,
\ee
where $\pi$ is a permutation of $\{2,3,\dots,n-1\}$ and $c_i$ are contracted dummy indices.\footnote{We suppress the dependence on momenta and polarization vectors for clarity.} Here, $b_{\a,\pi}\in\{0,\pm 1\}$ depends on the permutation $\pi$ \cite{Bern:2019prr} and on the index $\a$, which enumerates the Witten diagrams for given $n$ external legs \emph{with three-point vertices only}.\footnote{The colour factors of diagrams with four-point vertices coincide with those of diagrams with three-point vertices only, so the former can be absorbed into the latter; the existence of four-point vertices merely shifts the values of $W_\a$.}

For given $n$ external legs, we obtain the full amplitude by summing over all Witten diagrams:
\be
\label{eq: expansion of color dressed amplitude in KK basis}
\cA^{a_1\dots a_n}=&\sum\limits_{\a}\cW_{\a}^{a_1\dots a_n}
\\
=&\sum\limits_{\pi\in S_{n-2}}
\cA\left(1,\pi(2),\dots,\pi(n-1),n\right)
\\&\times f^{a_1a_{\pi(2)}c_1}\cdots f^{c_{n-3}a_{\pi(n-1)}a_n}\;,
\ee
where
\be
\label{eq: color ordered from Witten diagrams}
\cA\left(1,\pi(2),\dots,\pi(n-1),n\right)\equiv\sum\limits_{\a}W_{\a}\,b_{\a,\pi}
\ee
is the colour-ordered amplitude in the Kleiss--Kuijf (KK) basis. \eqref{eq: expansion of color dressed amplitude in KK basis} is usually referred to as the Del Duca--Dixon--Maltoni (DDM) decomposition of the colour-dressed amplitude \cite{DelDuca:1999iql,DelDuca:1999rs,Kleiss:1988ne,Bern:2019prr}.\footnote{Not all colour-ordered amplitudes are independent: in flat space, BCJ relations \cite{Bern:2008qj} relate them to one another. In AdS, the relation between colour-kinematics duality and amplitude relations is more subtle \cite{Albayrak:2020fyp}; BCJ-type relations involving differential operators or deformed kinematics were obtained in \cite{Armstrong:2020woi,Diwakar:2021juk,Zhou:2021gnu,Herderschee:2022ntr,Cheung:2022pdk}. }

The DDM decomposition is convenient as it involves only $(n-2)!$ terms. A more conventional basis is the trace basis,
\be
\cA^{a_1\dots a_n}
={}&\sum\limits_{\s\in S_{n-1}}
\cA\left(1,\s(2),\dots,\s(n)\right)
\\&\times\mathrm{Tr}\left(T^{a_1}T^{a_{\s(2)}}\cdots T^{a_{\s(n)}}\right)\;,
\ee
where $T^a$ are the generators of $\SU(N)$ in the fundamental representation, normalized as $\mathrm{Tr}(T^aT^b)=\de^{ab}$ and $[T^a,T^b]=i\sqrt2 f^{abc}T^c$, so that $f^{abc}=-\frac{i}{\sqrt2}\big[\mathrm{Tr}(T^aT^bT^c)-\mathrm{Tr}(T^aT^cT^b)\big]$. Using $\sum_a(T^a)_{i_1}^{j_1}(T^a)_{i_2}^{j_2}=\de_{i_1}^{j_2}\de_{i_2}^{j_1}-\de_{i_1}^{j_1}\de_{i_2}^{j_2}/N$, one can go back and forth between the DDM and trace bases. The partial amplitudes agree once $f^{abc}$ in \eqref{eq: expansion of color dressed amplitude in KK basis} is replaced by $\tilde f^{abc}\equiv i\sqrt2f^{abc}=\mathrm{Tr}([T^a,T^b]T^c)$. With $f^{abc}$ itself, the DDM coefficients are $(i\sqrt2)^{n-2}$ times the trace-basis partial amplitudes. The Kleiss--Kuijf relations map $(n-1)!$ partial amplitudes of trace basis to the $(n-2)!$ of DDM basis \cite{Kleiss:1988ne,Bern:2019prr}.

\subsection{Scalar seeds: slab integrals and Langevin diagrams}
\label{app: seeds}
This appendix collects the scalar ingredients of \S~{\ref{sec: flows}}: the slab propagators and the identities they obey, the closed forms of the exchange seed, and the Langevin diagrams whose sum gives the stochastic quadruple-trace weight. The seed theory is a massless scalar on the flat half-space with cubic coupling $\frac{\l}{3!}\f^3$, i.e.\ the conformally coupled $\f^3$ theory in AdS$_6$ after rescaling; its radial integrals do not depend on the boundary dimension. Its equilibrium action has $G^{(3)}=\l/(3k_{123})$ and, in the $s$-channel, $G^{(4)}_s=-\l^2/(12\,k_{1234}k_{12\underline{12}}k_{34\underline{12}})$, so that the slab coupling is $D^{(4)}_s=-\frac{\l^2}{24}F(\e)$ with $F$ as in \eqref{eq: F}.

\paragraph*{Slab propagators and bridge identities.} The bulk-to-cutoff and bulk-to-bulk propagators of the slab $0<z<\e$ are $S_k$ and $\cG^{\rm UV}_k$ of \eqref{eq: UV propagators}. Since $\cG^{\rm UV}_k$ is the Green's function of $-\partial_z^2+k^2$ with Dirichlet conditions at both ends,
\be
\label{eq: bridge identity}
(\partial_z+k)(\partial_w+k)\,\cG^{\rm UV}_k(z,w)=\de(z-w)-\frac{k\,e^{k(z+w-\e)}}{\sinh k\e}\;,
\ee
as follows from $(\partial_z+k)\sinh kz=k\,e^{kz}$ and $(\partial_w+k)\sinh k(\e-w)=-k\,e^{-k(\e-w)}$ together with the jump $-1$ of $\partial_z\cG^{\rm UV}_k$ across $z=w$. Its $k\to0$ limit is $\partial_z\partial_w\cG^{\rm UV}_0=\de(z-w)-1/\e$, the covariance identity of the Brownian bridge used in \eqref{eq: Coulomb zero mode radial}; at finite $k$, \eqref{eq: bridge identity} is used in \eqref{eq: Coulomb zero mode alpha} with $k\to\g$.

\paragraph*{Closed form of the exchange seed.} Expanding the products of $\sinh k_rz$ into hyperbolic cosines reduces $F(\e)$ to the function
\be
\cJ_s(a,b)\coloneqq\int_0^\e\!\!\int_0^\e dz\,dw\,\cosh(az)\,\cG^{\rm UV}_s(z,w)\cosh(bw)\;,
\ee
as in \eqref{eq: F closed}. The integrals are elementary and give
\bea[eq: J]
\cJ_s(a,b)&=\frac{X_s(a,b)}{(s^2-a^2)(s^2-b^2)}\;,\\
X_s(a,b)&=\frac{s\,(\mathrm{ch}_a+\mathrm{ch}_b-\mathrm{ch}_s-\mathrm{ch}_a\mathrm{ch}_b\mathrm{ch}_s)}{\mathrm{sh}_s}\nn\\
&\quad+\frac{(s^2-b^2)\,a\,\mathrm{sh}_a\mathrm{ch}_b-(s^2-a^2)\,b\,\mathrm{ch}_a\mathrm{sh}_b}{a^2-b^2}\;,
\eea
with $\mathrm{ch}_x\equiv\cosh x\e$ and $\mathrm{sh}_x\equiv\sinh x\e$. $\cJ_s$ is symmetric in $a\leftrightarrow b$ and even in each argument, and its apparent poles at $a^2=b^2$ and $s^2\in\{a^2,b^2\}$ are removable. The longitudinal seed $F^{(0)}=F|_{s\to0}$ follows from
\be
\label{eq: J0}
\cJ_0(a,b)={}&\frac{1}{a^2b^2}\bigg[\frac{\mathrm{ch}_a+\mathrm{ch}_b-1-\mathrm{ch}_a\mathrm{ch}_b}{\e}
\\&+\frac{ab\,(a\,\mathrm{ch}_a\mathrm{sh}_b-b\,\mathrm{sh}_a\mathrm{ch}_b)}{a^2-b^2}\bigg]\;.
\ee

\paragraph*{Langevin diagrams for the exchange identity.} On the Langevin side, $R_k(\tau)=e^{-k\tau}\theta(\tau)$ and $\cG_k(t,t')$ are as in \eqref{eq: scalar propagator}, and $g_k(t)=\cG_k(t,t)$. At order $\l^2$, the $s$-channel part of the connected stochastic four-point function is a sum of three terms. With $r\in\{1,2\}$, $r'\in\{3,4\}$, $\bar r$ the partner of $r$ in the channel, $\kappa\equiv k_{\underline{12}}$, and $G_3^L\equiv G^{(3)}(k_1,k_2,\kappa)$, $G_3^R\equiv G^{(3)}(\kappa,k_3,k_4)$:
\begin{subequations}
\label{eq: Langevin diagrams}
\begin{multline}
A=9\,G_3^LG_3^R\bigg[\sum_{r}\int_0^tds\,R_{k_r}(t-s)\,\cG_{k_{\bar r}}(s,t)\\
\x \int_0^sds'\,R_{\kappa}(s-s')\,\cG_{k_3}(s',t)\,\cG_{k_4}(s',t)\\+(12\leftrightarrow34)\bigg],
\end{multline}
\begin{multline}
B=9\,G_3^LG_3^R\sum_{r,r'}\int_0^t\!\!\int_0^tds\,ds'\,R_{k_r}(t-s)\\\x R_{k_{r'}}(t-s')\,\cG_{k_{\bar r}}(s,t)\,\cG_{\kappa}(s,s')\,\cG_{k_{\bar r'}}(s',t),
\end{multline}
\begin{multline}
C=-12\,G^{(4)}_s\sum_{r=1}^4\int_0^tds\,R_{k_r}(t-s)\prod_{r''\neq r}\cG_{k_{r''}}(s,t)\;.
\end{multline}
\end{subequations}
$A$ has a response function as internal line and arises from iterating the cubic drift; $B$ has a correlator as internal line; $C$ comes from the quartic term of $\SE$. Inserting $\<\f^4\>_c=A+B+C$ into \eqref{eq: P4 extraction} and comparing with $D^{(4)}_s=-\frac{\l^2}{24}F(t)$ gives $D^{(4)}_s=P^{(4)}_s-\half G^{(4)}_s$, the exchange identity behind \eqref{eq: P4}. Proposition~1 guarantees it; we have also verified it directly for generic norms and in the limits $\kappa\to0$ and $k_2\to0$ (see the ancillary file). A direct analytic evaluation of $A+B+C$, inserting \eqref{eq: I4} and \eqref{eq: J}--\eqref{eq: J0} together with the elementary time integrals of the response functions and correlators, would turn this into an algebraic identity; since Proposition~1 already establishes the result, we leave it to future work.

\subsection{WKB hierarchy for the radial ground state}
\label{app: transport}
For completeness we recall how the loop corrections to a ground-state wavefunction are determined by its tree-level part, and then record the form this takes for the kernels of $\SE$. Write $\PsiIR=e^{-\SE/2}$ with $\SE=\SE^{(0)}+\SE^{(1)}+\dots$ ordered by loops, and $H_{RG}=\int d^dx\,[-\half\de^2/\de\Phi^2+\cV]$. Since $-\half\nabla^2e^{-\SE/2}=[-\frac18(\nabla\SE)^2+\frac14\nabla^2\SE]\,e^{-\SE/2}$, the eigenvalue equation $H_{RG}\PsiIR=E_0\PsiIR$ reads
\be
\label{eq: WKB hierarchy}
\cV=\frac18(\nabla\SE)^2-\frac14\nabla^2\SE+E_0\;,
\ee
which is $H_{FP}=H_{RG}-E_0$ with $H_{FP}$ as in \eqref{eq: HFP}. At lowest order this is the Hamilton--Jacobi equation \eqref{eq: stationary HJ}, $\frac18(\nabla\SE^{(0)})^2=\cV$; at the next order it is the transport equation \eqref{eq: transport}. These are the first two equations of the WKB hierarchy of the Euclidean Schr\"odinger equation \cite{BerryMount:1972,MaslovFedoriuk:1981}, here for the functional Schr\"odinger equation \cite{Symanzik:1981wd,Jackiw:1990,Hatfield:1992} of the radial evolution \cite{Mansfield:1999kk}.

Equation \eqref{eq: transport} is linear and first order in $\SE^{(1)}$, so it is solved by the method of characteristics. The characteristics are the integral curves of $\nabla\SE^{(0)}$, i.e.\ the noiseless Langevin flow $\partial_z\Phi=-\half\nabla\SE^{(0)}$, which are the regular classical solutions of \S~{\ref{sec: regular branch}}. Along such a curve $\partial_z\SE^{(1)}[\Phi(z)]=-\half\nabla\SE^{(0)}\.\nabla\SE^{(1)}=-\half(\nabla^2\SE^{(0)}-4E_0)$, hence
\begin{multline}
\label{eq: S1 along flow}
\SE^{(1)}[\Phi]=\half\int_0^\infty dz\bigg[\int d^dx\,\frac{\de^2\SE^{(0)}}{\de\Phi(x)\,\de\Phi(x)}\bigg|_{\Phi_{\rm cl}(z)}-4E_0\bigg]\\+\SE^{(1)}[0]\;,
\end{multline}
where $\Phi_{\rm cl}(z)$ is the regular solution with $\Phi_{\rm cl}(0)=\Phi$. Convergence at the fixed point $\Phi_{\rm cl}(\infty)=0$ fixes $E_0=\frac14\nabla^2\SE^{(0)}[0]=\half\mathrm{Tr}\,\Omega$, the zero-point energy; for free fields $\SE^{(1)}$ is a constant. Equation \eqref{eq: S1 along flow} is the one-loop fluctuation determinant around the regular solution, written as a radial integral in the manner of Gel'fand--Yaglom \cite{Dunne:2007rt}.

Expanded in the kernels of \eqref{eq: multitrace dictionary}, with $\SE^{(L)}=\sum_nG^{(n,L)}\Phi^n$, the term of order $\Phi^n$ of \eqref{eq: transport} reads
\begin{multline}
\label{eq: transport kernels}
2k_{1\cdots n}\,G^{(n,1)}+\dots=(n+1)(n+2)\\\x\int_{\bm p}G^{(n+2,0)}(\bm k_1,\dots,\bm k_n,\bm p,-\bm p)
\end{multline}
for transverse legs, where the dots are products of tree kernels with lower-point one-loop kernels: the one-loop $n$-point kernel is the tree $(n+2)$-point kernel with two legs sewn, divided by the total energy of the external legs. For a quartic contact kernel, \eqref{eq: transport kernels} reproduces the one-loop tadpole $\int_{\bm p}\int_0^\infty dz\,e^{-2kz}\,\cG_p(z,z)=\int_{\bm p}\frac{1}{2k(2k+2p)}$ of the double-trace coupling, the factor $1/2p$ of the propagator cancelling against its $1-e^{-2pz}$. Equation \eqref{eq: transport kernels}, applied to the closed-form quartic kernel \eqref{eq: SE kernels} of Yang--Mills theory with the regularization of the sewn legs that point~(ii) of \S~{\ref{sec: outlook}} calls for, is the starting point of the one-loop computation in progress \cite{Albayrak:wip}; we do not pursue it here.

\bibliographystyle{utphysModified}
\bibliography{references}
\end{document}